\documentclass[%
 aip,
 jmp,%
 amsmath,amssymb,
 preprint,%
]{revtex4-1}

\usepackage{graphicx}
\usepackage{dcolumn}
\usepackage{bm}
\usepackage[off]{auto-pst-pdf}
\usepackage{psfrag}
\usepackage{bm}
\usepackage{subcaption}
\usepackage{array}
\usepackage{xr}
\newcolumntype{x}[1]{>{\centering\arraybackslash\hspace{0pt}}p{#1}}

\makeatletter
\renewcommand\@makecaption[2]{%
  \par
  \vskip\abovecaptionskip
  \begingroup
   \small\rmfamily
    \begingroup
     \samepage
     \flushing
     \let\footnote\@footnotemark@gobble
     \@make@capt@title{#1}{#2}\par
    \endgroup
  \endgroup
  \vskip\belowcaptionskip
}
\makeatother

\newcommand*{\citen}[1]{%
  \begingroup
    \romannumeral-`\x 
    \setcitestyle{numbers}%
    \cite{#1}%
  \endgroup   
}

\begin{document}


\title[]{Archetypal solution spaces for clustering gene expression datasets in identification of cancer subtypes} 

\author{Y. Wu}
\affiliation{ 
Yusuf Hamied Department of Chemistry, Lensfield Road, Cambridge CB2 1EW, United Kingdom
}
\author{L. Dicks}
\affiliation{ 
Yusuf Hamied Department of Chemistry, Lensfield Road, Cambridge CB2 1EW, United Kingdom
}
\author{D. J. Wales}%
 \email{dw34@cam.ac.uk}
\affiliation{ 
Yusuf Hamied Department of Chemistry, Lensfield Road, Cambridge CB2 1EW, United Kingdom
}

\date{\today}

\begin{abstract}
Gene expression profiles are essential in identifying different cancer
phenotypes. Clustering gene expression datasets can provide accurate
identification of cancerous cell lines, but this task is challenging due to the
small sample size and high dimensionality. Using the $K$-means clustering
algorithm we determine the organisation of the solution space for a variety
of gene expression datasets using energy landscape theory. The solution space landscapes allow us to understand $K$-means performance, and guide more effective use when varying common dataset properties; number of features, number of clusters, and cluster distribution. We find that the landscapes have a single-funnelled structure for the appropriate number of clusters, which is lost when the number of clusters deviates from this. We quantify this landscape structure using a frustration metric and show that it may provide a novel diagnostic tool for the appropriate number of cancer subtypes.
\end{abstract}


\maketitle

\section{Introduction}

The development of DNA microarray technologies has enabled the identification of differentially expressed genes in cells.\cite{TusherTC01} Variation in gene expression reflects changes in biological processes and cell environment, and clustering to identify cancer tissue samples sharing similar gene expression patterns has been used to discover new cancer subtypes.\cite{GolubST99, Jiang2004} An accurate cancer classification can be used to identify key genes\cite{Bae2021, Habib2022, Kothari2020} and develop models for accurate classification of new patients into the cancer subtypes.\cite{Yogita2022, Liu2022, Shen2022, Gao2019} Hence, the clustering quality is of central importance to facilitating the development and application of precision medicine, which has greater efficacy and specificity against a targeted tumour.

Clustering the gene expression data of cancer tissue samples remains challenging because these datasets have a small number of high-dimensional samples. Each feature is the expression of a given gene, and there are typically of $\mathcal{O}(10^4)$. Clustering in high dimensions is challenging,\cite{Koepke2010, Klawonn2015} due in part to the resulting uniformity of distances.\cite{Beyer1999, Jaskowiak2014} Therefore the dimensionality is usually reduced and clustering performed in a subspace,\cite{Parsons2004, Vidal2011} although this approach can be counter-productive for gene expression data.\cite{Yeung2001}

Many different clustering algorithms have been used for cancer gene expression datasets,\cite{Dhaeseleer05, OyeladeIO16, DeSouto08} for example hierarchical clustering,\cite{Sorlie2001, Sotirious2003, ElRehim2005} matrix factorisation,\cite{Brunet2004} concensus clustering,\cite{Monti2003, DeSousa2013, Wang2013} integrative clustering,\cite{Curtis2012} similarity network fusion\cite{Xu2016} and spectral clustering.\cite{Jiang2019} A popular choice is $K$-means,\cite{Macqueen67} which has been widely applied to gene expression data, along with its variants.\cite{Martella2012, Yao2016, Sasic2017, Kiselev2017, Jothi2019, Guo2019, Khan2021, Huang2021, Singh2022} In addition to a good separation between cancer subtypes at a chosen number of clusters it is important to select the appropriate number of clusters, which is not known \textit{a priori}. Estimating the number of clusters can be a challenging task and there can be variation in the predicted number of cancer subtypes.\cite{Dai2015} Most methods rely on metrics, such as the Dunn index,\cite{Dunn1973} Davies-Bouldin index\cite{Davies1979} or the silhouette coefficient,\cite{Rousseeuw1987} which estimate the quality of clustering at a given cluster number based on the variance of, and separation between, clusters. Clustering is performed at a range of cluster numbers and the optimal number is selected based on the index evaluated for a single clustering at each cluster number.

The potential energy surface is the fundamental object of interest in molecular science, from which all thermodynamic and kinetic information can be extracted. Various tools have been developed to explore and analyse these high-dimensional surfaces. The $K$-means cost function surface is analogous to the potential energy surface, and the computational tools developed in the energy landscape framework\cite{Wales03} can be adapted to identify minima and transition states on the $K$-means cost function surface.\cite{DicksW22} In this contribution, we construct $K$-means cost function landscapes for cancer gene expression datasets and analyse their structure to provide new insight into the success or failure of $K$-means clustering for gene expression data.

The paper is organised as follows. First, we present the application of the energy landscape framework to the $K$-means cost function, to elucidate the structure of the solution landscapes. In Section III we present the solution landscapes for gene expression datasets as we vary the clinical number of clusters, number of features, and cluster distribution. In the final section we analyse the changes in solution space as a result of these dataset features, and discuss the implications for clustering to determine cancer subtypes.

\section{Methods}

\subsection{Gene expression datasets}

In this work, we analyse the microarray gene expression datasets curated by de Souto \textit{et~al.}~[\citen{DeSouto08}], as summarised in
Table~\ref{tab:datasets}. The original datasets contain $\mathcal{O}(10^4)$ genes before dimensionality reduction to filter out uninformative genes, which is described in the original publication. Dimensionality reduction is essential to remove redundant genes and allow for more accurate clustering, and resulted in $\mathcal{O}(10^3)$ genes after processing. 

There are many other gene expression datasets available at the Cancer Genome Atlas (TCGA). The datasets selected for use in our work are of typical size and number of cancer subtypes, and provide the variety to probe general features of importance to clustering performance: number of clusters, number of features, and cluster distribution. The raw gene expression values have significantly different ranges, which would give features with larger values more influence during $K$-means clustering. Various standardisations were examined in [\citen{DeSouto08}], and we apply MinMax normalisation to scale each feature within the range [0, 1] throughout this work. Standardisation will affect the structure of $K$-means landscapes, but our focus was the effect of dataset properties.

\begin{table}
\caption[Information about gene expression datasets]{Summary of gene
expression datasets from de Souto \textit{et~al.}~[\citen{DeSouto08}]. The number of genes presented is after dimensionality reduction.}
\centering
\label{tab:datasets}
\begin{tabular}{x{3.5cm} x{2.5cm} x{1.75cm} x{1.75cm} x{3.5cm} x{1.75cm} }
\hline 
Name & Abbreviation & Number of samples & Number of classes & Samples per class & Number of genes \\ 
\hline
Yeoh-2002-v1 & Yeov1 & 248 & 2  & 43, 205 & 2526\\
Yeoh-2002-v2 & Yeov2 & 248 & 6 & 43, 15, 27, 64, 20, 79 & 2526\\
Dyrskjot-2003 & Dyr & 40 & 3 & 9, 20, 11 & 1203\\

Tomlins-2006-v2 & Tomv2 & 92 & 4 & 27, 20, 32, 13 & 1288 \\

Singh-2002 & Sin & 102 & 2 & 50, 52 & 339\\

Alizadeh-2000-v1 & Aliv1 & 42 & 2 & 21, 21 & 1095\\

Bittner-2000 & Bit & 38 & 2 & 19, 19 & 2201 \\
Armstrong-2002-v1 & Armv1 & 42 & 2 & 24, 48 & 1081\\

Shipp-2002-v1 & Shiv1 & 77 & 2 & 19, 58 & 798\\

Gordon-2002 & Gor & 181 & 2 & 31, 150 & 1626\\
Nutt-2003-v3 & Nutv3 & 22 & 2 & 7, 15 & 1152 \\

\hline 
\end{tabular}
\end{table}

\subsection{$K$-means landscapes}

The procedures for constructing and visualising $K$-means landscapes were adapted from Ref.~[\citen{DicksW22}], as summarised below.

\subsubsection{$K$-means cost function}

For a given number of clusters, $K$, and initial cluster positions, $\bm{\mu}$, a $K$-means solution is obtained by minimising the sum-of-squares cost function,
\begin{equation}  \label{eqn:cost}
J(\bm{\mu})=\sum_{i=1}^{N} \sum_{k=1}^{K} r_{ik} ||\bm{x}_i - \bm{\mu}_k ||^2 ,
\end{equation}
until a local minimum is attained. $N$ is the number of data points, each with $N_\mathrm{f}$ features, in the dataset $\bm{x}$. Each $\bm{x}_i$ is the position of data point $i$, and $\bm{\mu}_k$ the position of cluster $k$. $\bm{r}$ is an ($N \times K$)-dimensional matrix, which indicates the assignment of each data point to the $K$ clusters. Each data point is assigned to the nearest cluster in Euclidean distance, according to:
\begin{equation}
r_{ik} = 
\begin{cases}
1, & \text{if } k = \min\limits_j || \bm{x}_i - \bm{\mu}_{j} ||^2 ,\\
0, & \text{if otherwise.}
\end{cases}
\end{equation}

Starting from a specified number of clusters and given cluster positions, Lloyd's algorithm~\cite{Lloyd82} was used for local minimisation of $J$. Lloyd's algorithm iteratively assigns data points to their closest cluster centres and recalculates the centroids of each partition. The minimisation was considered converged when the cluster assignment remained unchanged between consecutive iterations.

The number of local minima of $J$, which correspond to clusterings of varying quality, can be vast. Consequently, the quality of $K$-means solutions can be highly dependent on the scheme for generating initial cluster positions, and location of the optimal solution is an NP-hard problem.\cite{Aloise2009, Mahajan2009} Many initialisation schemes have been developed,\cite{Arthur2007, Qi2017, Bachem2016} with variations for the large datasets that are becoming commonplace.\cite{Cohen2015, Sculley2010, Capo2017} In this study, we initialised cluster centres at $K$ randomly selected data points.\cite{Macqueen67}

Crucial to locating good $K$-means solutions, for all initialisation schemes, is the topography of the cost function surface. The surface topography, described by the $K$-means minima and the transition states that encode intermediate behaviour of the cost function, is controlled by both the cluster number and the dataset. Location of those minima with low values of the cost function, amongst the vast number of local minima, depends upon the organisation of the cost function surface.

\subsubsection{Characterising stationary points and pathways} \label{sec:locate_ts}

A transition state is defined as a point with zero gradient and a single negative eigenvalue of the Hessian matrix.\cite{murrelll68} On the $K$-means cost function surface, each possible cluster assignment $\bm{r}$ is associated with a quadratic, which is described by the double summation in Equation~\ref{eqn:cost}, and the cluster assignment changes at the intersection of two quadratics. 
Both the gradient and Hessian are undefined at a change in cluster assignment, so the cost function surface does not contain conventional transition states.
Instead, we can use the minimum of an intersection seam, known as the minimum-energy crossing point (MECP). An MECP that is a maximum orthogonal to the intersection seam is considered a transition state, and gives the minimum height of the cost function barrier between two different cluster assignments. 

To search for transition states between a chosen pair of minima, we construct an initial linear path between them. The path is composed of equally-spaced cluster positions, which are referred to as images. At any change in cluster assignment between two adjacent images, the first image is selected as a transition state candidate. To locate transition states from these candidates, a penalty-constrained MECP optimization algorithm~\cite{LevineCM08} was used.
For any pair of quadratics $J^{(1)} = J(\bm{\mu}, \bm{r}^{(1)})$ and
$J^{(2)} = J(\bm{\mu}, \bm{r}^{(2)})$, defined by the differing cluster assignments of adjacent images $\bm{r}^{(1)}$ and $\bm{r}^{(2)}$, 
the MECP between them can be found by minimising the surrogate function:
\begin{equation}
F_+(\bm{\mu}, \bm{r}^{(1)}, \bm{r}^{(2)}, \sigma, \alpha) = \frac{1}{2} \left( J^{(1)}+J^{(2)} \right) + 
\sigma \left( \frac{\left( J^{(1)} - J^{(2)} \right)^2}{ ||J^{(1)} - J^{(2)}|| + \alpha} \right) ,
\end{equation}
starting from the candidate cluster position. $\alpha$ is a smoothing parameter to prevent a discontinuity at the intersection seam, and $\sigma$ determines the penalty for deviation from the intersection seam. The choice of parameters is detailed in the SI.
The minimisation of $F_+$ is performed using the L-BFGS (limited-memory Broyden, Fletcher, Goldfarb, Shanno) algorithm.\cite{Nocedal80, LiuN89}
Only MECPs whose optimal cluster assignment is the same as either $\bm{r}^{(1)}$ or $\bm{r}^{(2)}$ are retained.
The cost function of the transition state between $\bm{r}^{(1)}$ and $\bm{r}^{(2)}$ is given by the value of $J$ at the MECP.

Conventionally, the two minima connected by a transition state can be found by following (approximate) steepest-descent paths parallel and antiparallel to the eigenvector corresponding to the single negative Hessian eigenvalue. 
The Hessian is undefined at the transition state analogue, so we construct another surrogate function to obtain the required eigenvector:
\begin{equation}
F_-(\bm{\mu}, \bm{r}^{(1)}, \bm{r}^{(2)}, \sigma, \alpha) = \frac{1}{2} \left( J^{(1)}+J^{(2)} \right) - 
\sigma \left( \frac{\left( J^{(1)} - J^{(2)} \right)^2}{ ||J^{(1)} - J^{(2)}|| + \alpha} \right).
\end{equation}
$\bm{r}^{(1)}$, $\bm{r}^{(2)}$, $\sigma$, and $\alpha$ are the same as in $F_+$.
Unlike $F_+$, $F_-$ penalises the two quadratics having similar values.
We calculate the Hessian of $F_-$ at the minimum of $F_+$ to find the negative eigenvalue and the corresponding eigenvector. 
Finally, the L-BFGS algorithm is used to calculate the approximate steepest-descent paths from the transition state analogue.

\subsubsection{Constructing the stationary point network}

Minima and transition states on the $K$-means cost function surface are added to a stationary point database. This database defines the $K$-means landscape and can be represented as a weighted graph in which each minimum is a node and there is an edge between two minima directly connected by a transition state. Such a graph is analogous to a kinetic transition network in molecular science.\cite{Wales09, Wales06, NoeF08} 

\begin{figure}[h!]
    \centering
    \includegraphics[width=1.0\textwidth]{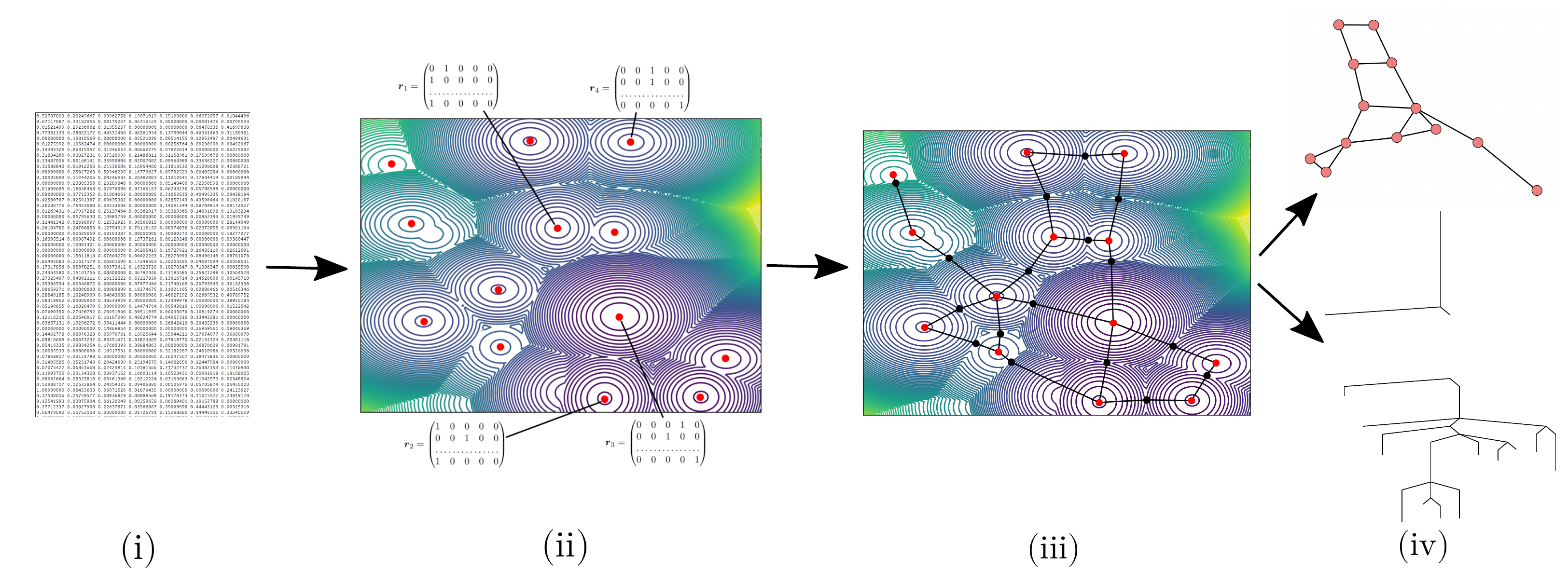}
    \caption{The computational workflow for generating $K$-means landscapes of gene expression data. (i) We begin from an unlabelled gene expression dataset and a specified $K$. (ii) Generate clustering solutions, each of which is a minimum on the cost function surface. Here, solutions are denoted by red circles with the corresponding cluster assignment matrices shown for selected minima. The cost function surface cannot be directly visualised for all but trivial cases, as its dimensionality is $K \times N$. (iii) Locate the transition states (black circles) and their connected minima, which are joined by solid black lines. (iv) We can represent this connected set of clustering solutions using both a weighted graph (top) and disconnectivity graphs (bottom).}
    \label{fig:workflow}
\end{figure}

We systematically select pairs of minima from the database and apply the algorithms in Sec.~\ref{sec:locate_ts} to locate intermediate transition states and minima. Pairs are selected to efficiently produce a connected set, where each minimum can reach any other through a sequence of transition states and connected minima. The selection of pairs for connection uses a distance-based criterion.\cite{RoederW18} The lowest-valued minimum that is not connected to the global minimum is paired with the nearest member in Euclidean distance that is connected. Any transition states, together with their connected minima, are added to the stationary point database if they are not already present. This criterion is applied repeatedly to select pairs until all minima are connected. An overview of the computational workflow is given in Fig.~\ref{fig:workflow}.

We use disconnectivity graphs\cite{beckerk97,walesmw98} to visualise the high-dimensional cost function surfaces, which include all clustering solutions and provide a faithful description of the barriers between them. For a given landscape, the minima can be separated into disjoint sets (referred to as superbasins~\cite{beckerk97}) at any threshold value for $J$. Two minima belong to different superbasins if the lowest barrier between them is greater than this threshold. The pathway between a pair of minima may involve a sequence of connected minima and transition states, and the overall barrier between them is given by $J$ at the highest transition state along the pathway. To construct the disconnectivity graph, the minima are separated into superbasins at regular thresholds. The superbasins are represented as nodes arranged horizontally on the graph. The horizontal spacing is adjusted to arrange the nodes as clearly as possible, and the vertical axis corresponds to the cost function.
A vertical line extends upwards from each minimum, and any lines merge at the threshold when they belong to the same superbasin.

Disconnectivity graphs reveal the organisation of the landscape, and consequently the general characteristics of a physical system. In particular, the landscape may be single funnelled\cite{socciow98} or multi-funnelled.\cite{ChebaroBCW15}  Single-funnel landscapes are typical of self-organising systems, such as evolved proteins and have small barriers to lower-valued minima, leading to efficient relaxation to the global minimum. Multi-funnelled landscapes contain additional funnels that do not contain the global minimum, each of which can slow relaxation to the global minimum. These two paradigms are useful references for surface organisation.

\section{Results}

$K$-means landscapes for several gene expression datasets are presented in this section. These datasets are selected to examine the influence of several properties on the space of clustering solutions, namely the clinical number of clusters, number of features, and the cluster distribution. The landscapes are constructed without reference to the clinical labelling, although in Secs.~\ref{sec:variation of the number of clusters}--\ref{sec:varying_samples} we select $K$ to match the clinical number of clusters.

Throughout it is important to distinguish the quality of clustering solutions and their ease of location, both of which contribute to $K$-means performance. The quality of clustering can be measured with or without reference labellings using standard metrics, \textit{e.g.} Rand index,\cite{Rand1971} silhouette coefficient,\cite{Rousseeuw1987} of the Davies-Bouldin index.\cite{Davies1979} However, the topography of the solution space can provide unique information about the difficulty of the optimisation problem, and the robustness, reproducibility, and importance of locating low-valued solutions.

\subsection{Varying the number of clusters}
\label{sec:variation of the number of clusters}

\begin{figure}[h!]
    \centering
    \begin{subfigure}{0.3\textwidth}
        \centering
        \includegraphics[width=1.0\textwidth]{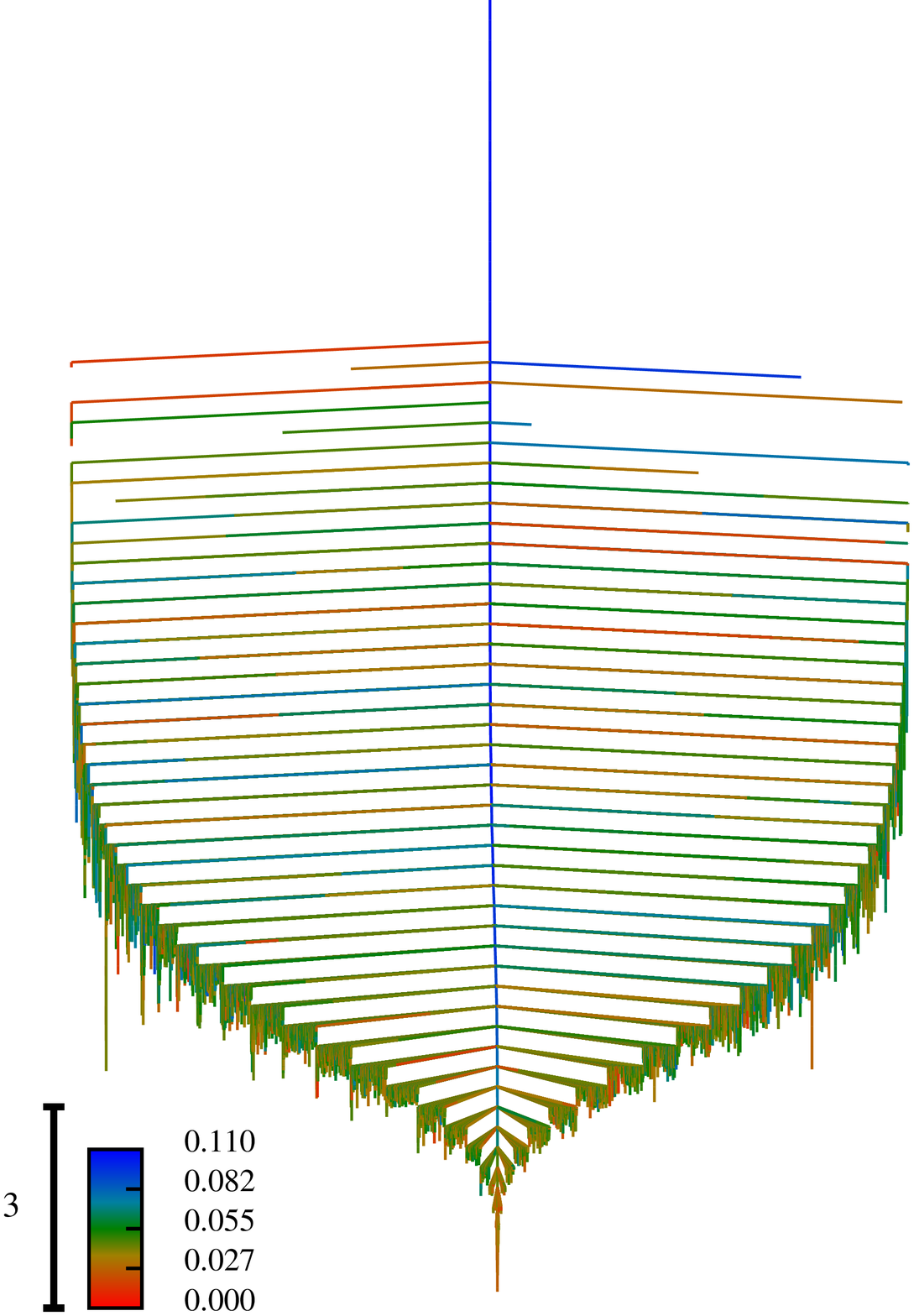}
        \caption{Yeov1 ($K=2$)}
    \end{subfigure}
    \begin{subfigure}{0.3\textwidth}
        \centering
        \includegraphics[width=1.0\textwidth]{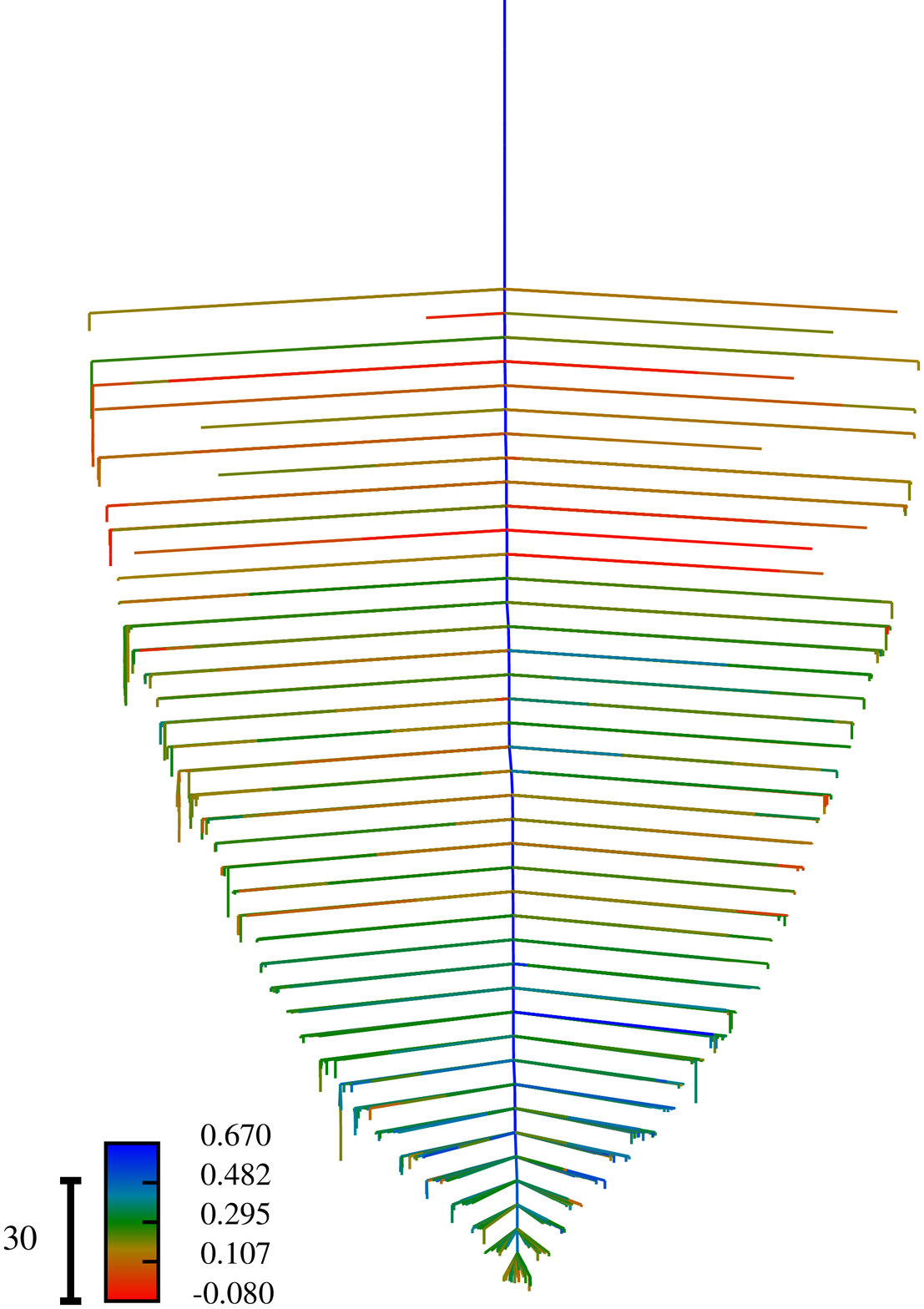}
        \caption{Dyr ($K=3$)}
    \end{subfigure}
    \begin{subfigure}{0.3\textwidth}
        \centering
        \includegraphics[width=1.0\textwidth]{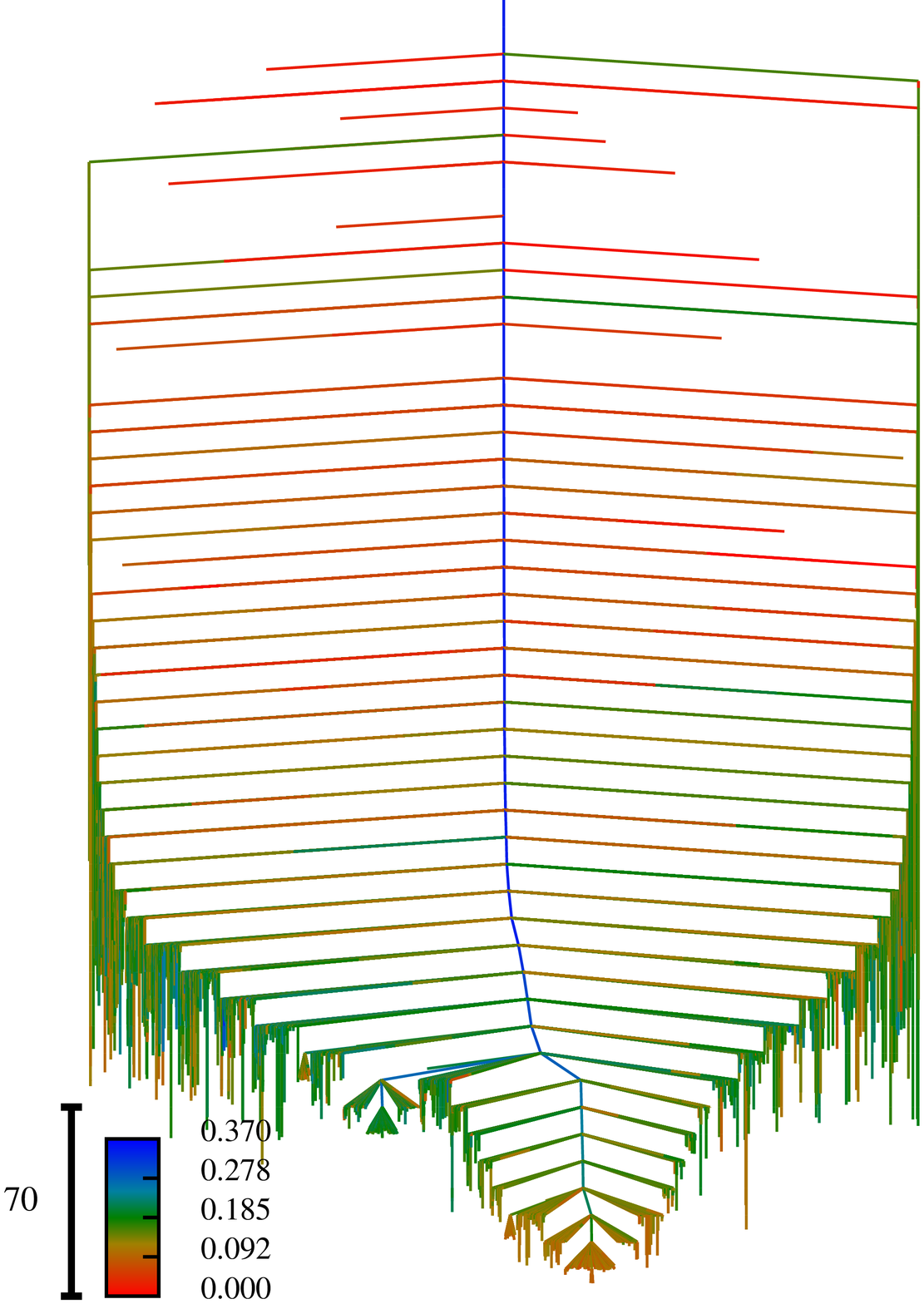}
        \caption{Tomv2 ($K=4$)}
    \end{subfigure}\\
    \caption[$K$-means landscapes for varying $K$]{Disconnectivity graphs for $K$-means landscapes with the Yeov1, Dyr and Tomv2 gene expression datasets. The vertical axis gives the value of the cost function. Each minimum on the cost function surface (distinct clustering solution) is denoted by a vertical line that begins at its corresponding cost function value. Lines merge at the cost function value at which they can be connected, and they are organised in the horizontal axis to aid visualisation. Each minimum is coloured by its adjusted Rand index relative to the specialist-determined true labels, whose value is shown in the colour bar. The scale bar denotes the cost function range.}
    \label{fig:trees_vary_k}
\end{figure}

Clustering generally becomes more challenging as the number of clusters increases. Here, we select three representative datasets (Yeov1, Dyr and Tomv2) with clinical assignments of 2, 3 and 4 clusters, respectively. In Figure~\ref{fig:trees_vary_k} we present the first visualisations of clustering solutions spaces for gene expression data and show that all landscapes have a largely funnelled structure despite the changing $K$. 

Single-funnelled landscapes have small barriers to lower-valued minima, which is indicative of efficient relaxation to the global minimum in molecular systems, and faster global optimisation. The solution landscape contains many shallow minima above the global minimum that have a small volume in configuration space and exist within a larger downhill basin. The small barriers ensure that new minima can be reached with little perturbation in cluster positions, and the minima higher in the landscape are therefore less robust. A small change in cluster coordinates, or dataset coordinates, can remove the minimum by subsuming it into another basin of attraction. However, because of this property the global minimum is more robust and reproducible as it can be reached by small perturbations from the less stable minima higher in cost function. Such single-funnel landscapes appear regularly in clustering gene expression data.

When $K$ increases to 4, the landscape contain a small subfunnel. Subfunnels are regions of the solution space that are locally funnelled to a minimum higher in energy than the global minimum. The minima within subfunnels are non-optimal alternative solutions. Once within a subfunnel there may be a significant barrier to the global minimum, which can impede further relaxation to the global minimum and degrade the performance of optimisation algorithms. However, if small barriers separate the subfunnel from the global minimum, relaxation remains efficient, as for this example with $K=4$.

\begin{figure}[h!]
    \centering
    \includegraphics[width=1.0\textwidth]{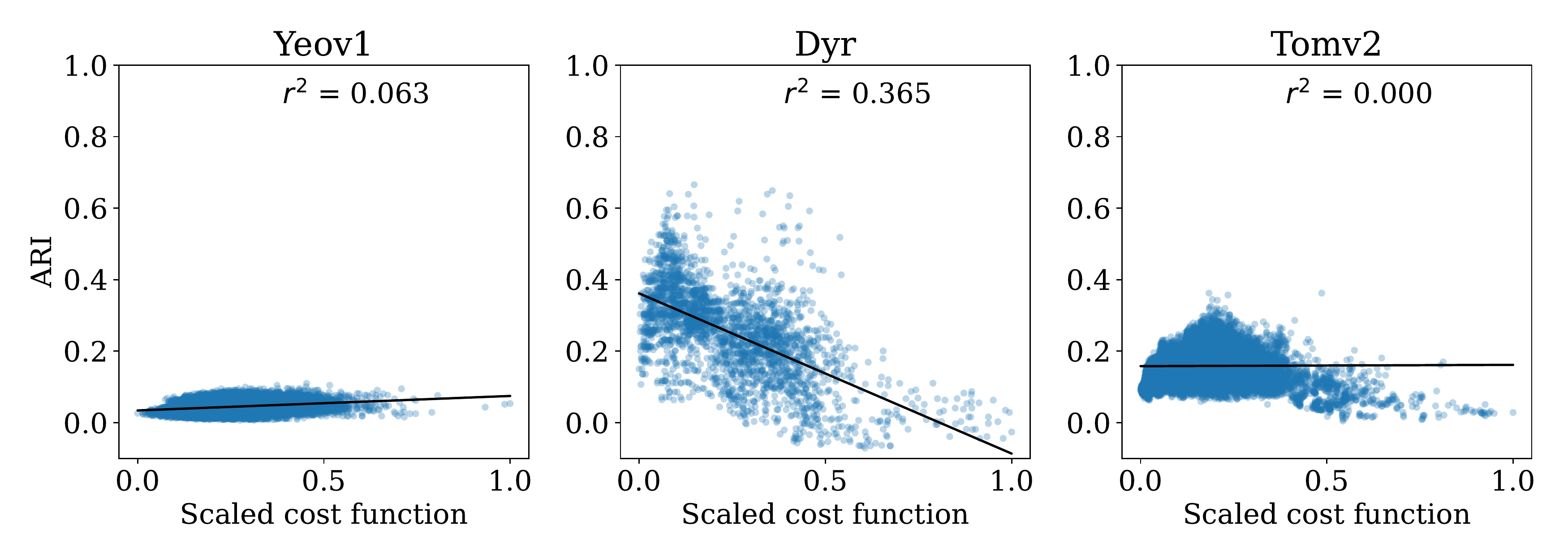}
     \caption[Histograms of ARIs of minima]{ Scatter plots of ARI relative to the specialist-determined true labels against the cost function for three gene expression datasets of different cluster number. The cost function is scaled to the range (0, 1).}
    \label{fig:dist_ri}
\end{figure}

To evaluate the accuracy of clustering solutions we calculate the adjusted Rand index (ARI)\cite{Rand1971} of each minimum relative to the clinical assignment. The ARI ranges from $-1$ to $+1$, and is higher when the clustering solutions are more similar. Figure~\ref{fig:dist_ri} shows that, for Yeov1 and Tomv2, most minima have ARIs close to zero, and these solutions are only slightly more accurate than random labelling. For the Dyr dataset the accuracy of $K$-means clustering is greater, with a median ARI of 0.26, and a maximum ARI of 0.67.

For the Dyr dataset, the most accurate clustering does not correspond to the global minimum of $J$. Instead, it lies at a height of 0.148 within the scaled cost function range. However, there is a correlation between cost function and cluster accuracy, indicating that locating minima lower in cost function will generally produce clusterings with better discrimination between cancer subtypes. For the two datasets with poor $K$-means performance there is no correlation between the ARI and the cost function value, due to the poor quality of all the clusterings. Therefore, better optimisation may not necessarily lead to improved clustering performance.

\begin{figure}[h!]
    \centering
    \includegraphics[width=1.0\textwidth]{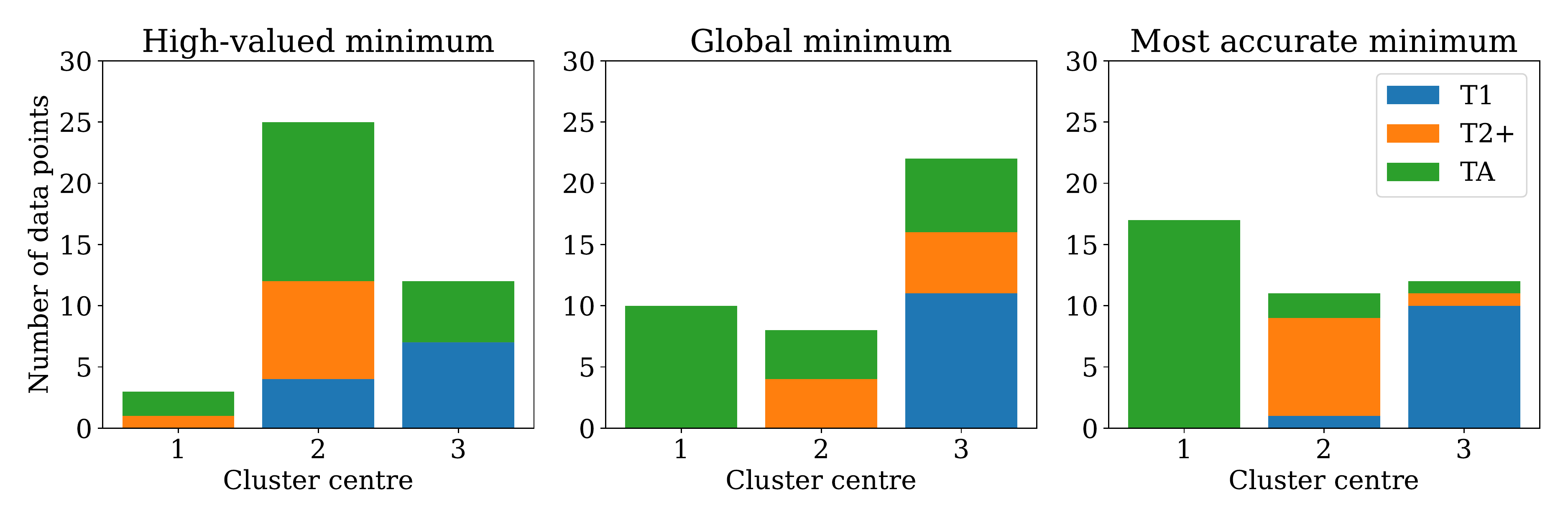}
     \caption[Composition of clustering solutions in the Dyr dataset]{Composition of selected clustering solutions of the Dyr dataset. T2+, TA and T1 are three subtypes of bladder carcinoma. 
A minimum from the high-cost function region (ARI = 0.058), the global minimum (ARI = 0.150) and the most accurate minimum (ARI = 0.665) are compared.}
    \label{fig:compos_dyr}
\end{figure}

The accuracy of $K$-means solutions can be understood through the cluster composition. We analyse the composition of three minima for the Dyr dataset: the global minimum, the most accurate $K$-means clustering, and a high-valued solution, as shown in Figure~\ref{fig:compos_dyr}. The minimum with
high $J$ exhibits poor accuracy due to a mixture of the clinical cancer subtypes T2+, TA and T1 across the $K$-means clusters. The global minimum clustering achieves a better separation of data points and successfully identifies a cluster that contains only data points labelled `TA'. However, we see that the most accurate clustering achieves an even better separation of data points than the global minimum, and all three classes are nearly separated.

Therefore, the solution that minimises the $K$-means cost function leads to a poorer separation of the T1 and T2+ cancer subtypes. Furthermore, the most accurate clustering has good performance, but the ARI fluctuates widely within the low cost function region, indicating that a set of low-valued minima may give very different results. The large number of minima that have poor performance for all datasets highlight the challenge of clustering gene expression datasets. The optimisation problem of locating the global minimum remains similar in all funnelled landscapes, as the appropriate number of clusters appears not to change the organisation of solution space. However, the cost function may be an inappropriate surrogate for accuracy for high-dimensional gene expression datasets with few data points, leading to the need for ensemble methods or external validation metrics to avoid poor performance.

\subsection{Varying the number of features}
\label{sec:vary_nf}

As the number of features increases, $K$-means clustering generally becomes more challenging. The distances between data points become more uniform\cite{Klawonn2015} and discrimination between the nearest and farthest points becomes more difficult. Here, we select three representative gene expression datasets (Sin, Aliv1 and Bit) that have a different number of features (339, 1095 and 2201, respectively), but the same clinical number of clusters ($K=2$), and similar samples per class ratios (approximately $1:1$).

\begin{figure}[h!]
    \centering
    \begin{subfigure}{0.3\textwidth}
        \centering
        \includegraphics[width=1.0\textwidth]{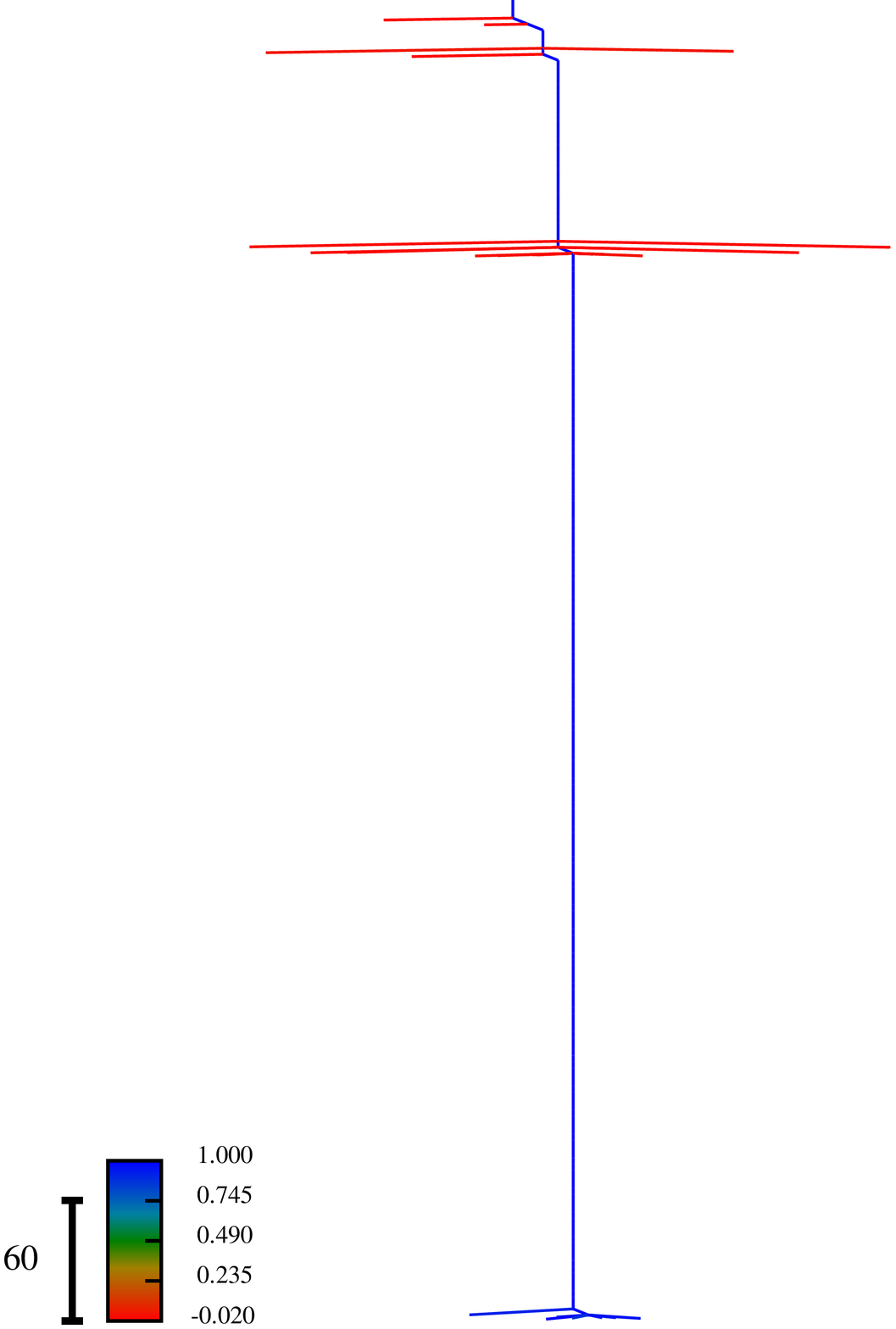}
        \caption{Sin ($N_\mathrm{f}=339$)}
    \end{subfigure}%
    \begin{subfigure}{0.3\textwidth}
        \centering
        \includegraphics[width=1.0\textwidth]{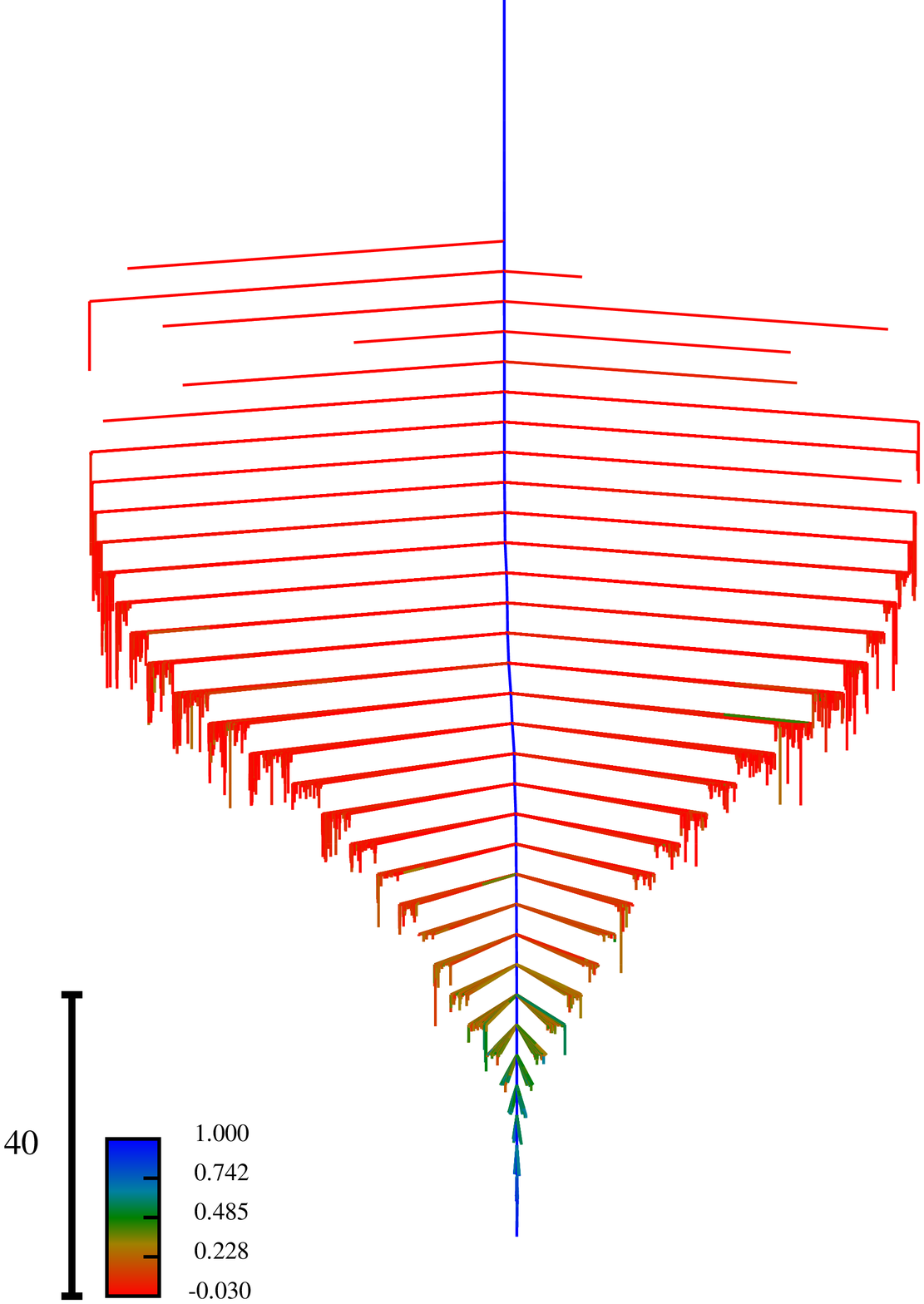}
        \caption{Aliv1 ($N_\mathrm{f}=1095$)}
    \end{subfigure}%
    \begin{subfigure}{0.3\textwidth}
        \centering
        \includegraphics[width=1.0\textwidth]{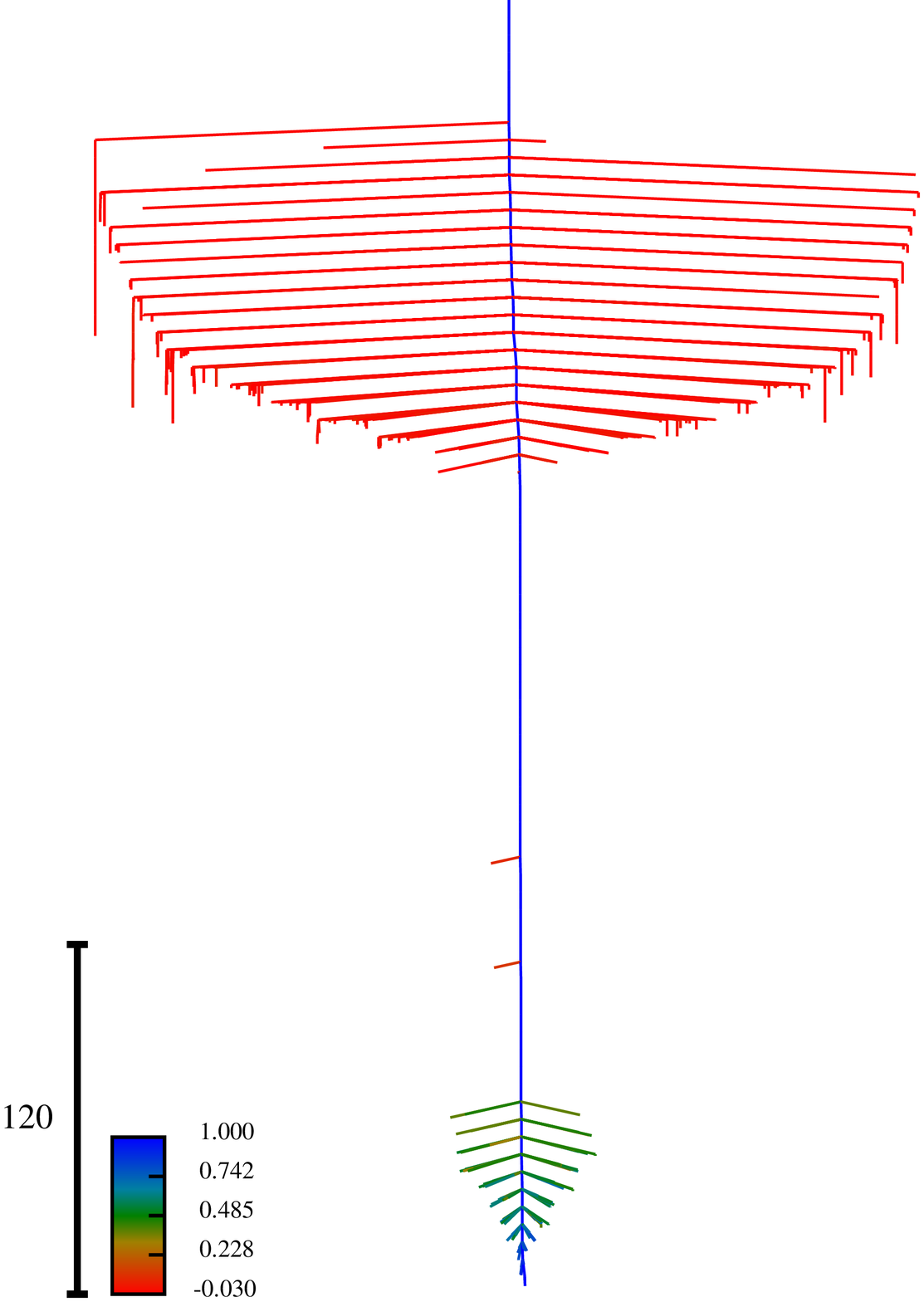}
        \caption{Bit ($N_\mathrm{f}=2201$)}
    \end{subfigure}\\
    \caption[$K$-means landscapes for a varying number of features]{$K$-means landscapes for gene expression datasets with a varying number of features. $K=2$ for all datasets. Minima are coloured by their ARIs relative to the global minimum.
The range of ARIs is shown in the colour bar. 
The scale bar denotes the cost function value.}
    \label{fig:trees_vary_features}
\end{figure}

The increase in the number of features leads to an increase in the number of minima, but does not affect the overall single-funnel structure of the landscapes, as shown in Figure~\ref{fig:trees_vary_features}. There remains a funnelled organisation of the solution space, which supports straightforward relaxation to the global minimum. The ability of $K$-means to accurately reproduce the true clinical labels differs between datasets (maximum ARI = 0.068, 0.905 and 0.613 for Sin, Aliv1 and Bit datasets respectively), but the optimisation problem remains of similar complexity.

Despite the increased number of minima the organisation of the landscapes shows that there is still a single dominant clustering motif that can be easily reached from all other clustering solutions via low barrier pathways. The number of minima resembling this global minimum clustering increases with the number of features, which reduces the importance of locating the true global minimum, as there are a variety of comparable clusterings at similar $J$.

For the Sin and Bit datasets, there are significant gaps along the vertical axis of the disconnectivity graphs. The high-lying minima contain cluster centres placed between outliers and non-outliers, which are unstable and have a small cost function barrier to lower-valued minima. These solutions can be relaxed to the global minimum region by displacing the cluster centres away from a small number of outliers, and minimising, which separates assignment of outliers and non-outliers into different clusters.

\subsection{Varying the samples per class ratio} \label{sec:varying_samples}

The samples per class ratio measures the distribution of cluster sizes in the dataset. Uneven distributions will make clustering more challenging because small clusters may not be easily distinguished. Here, we select four representative gene expression datasets (Bit, Armv1, Shiv1 and Gor) with different samples per class ratios (approximately $1:1$, $1:2$, $1:3$ and $1:5$), but the same number of classes ($K=2$).

\begin{figure}[h!]
    \centering
    \begin{subfigure}{0.25\textwidth}
        \centering
        \includegraphics[width=0.95\textwidth]{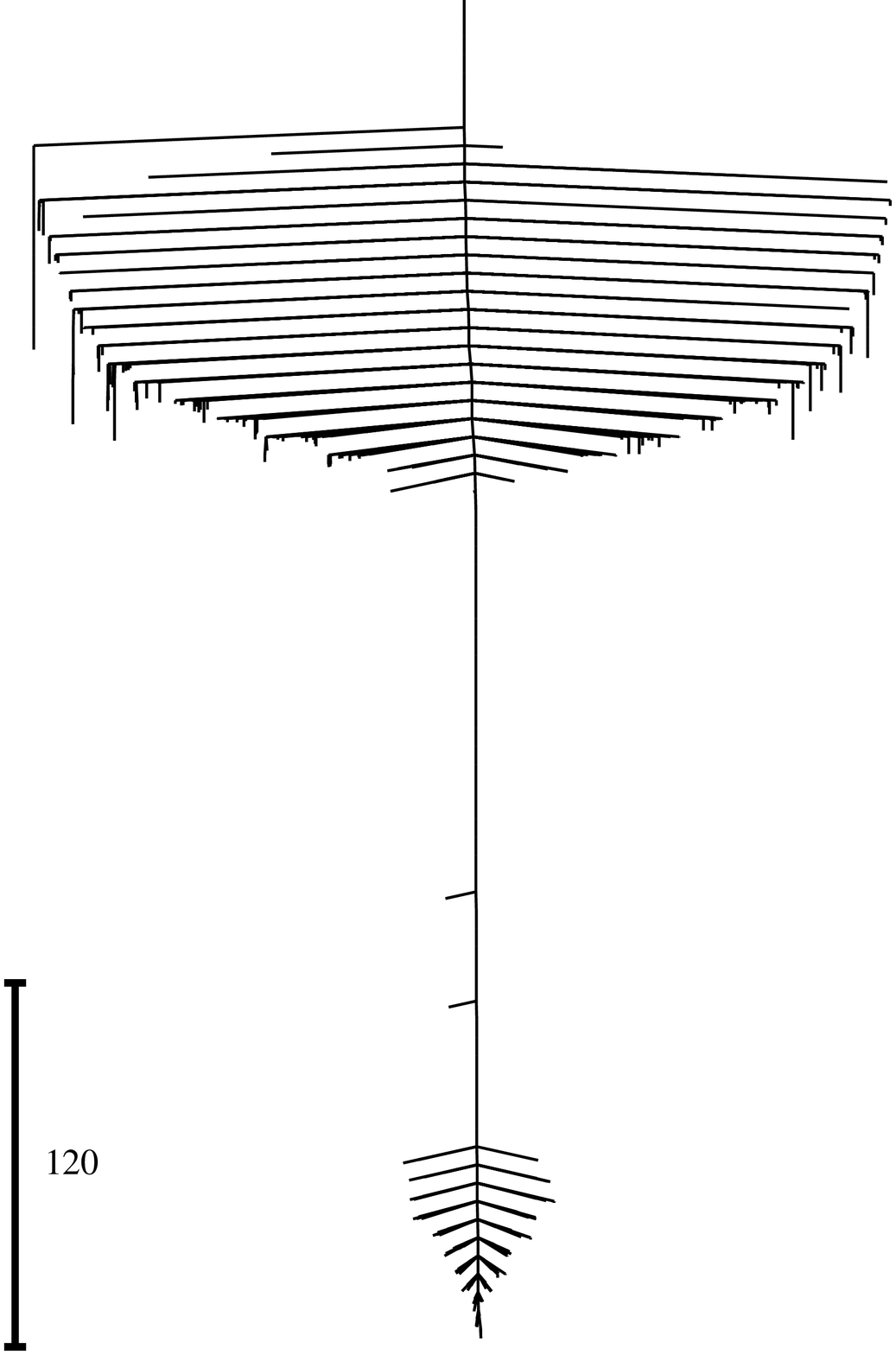}
        \caption{Bit ($1:1$)}
    \end{subfigure}%
    \begin{subfigure}{0.25\textwidth}
        \centering
        \includegraphics[width=0.95\textwidth]{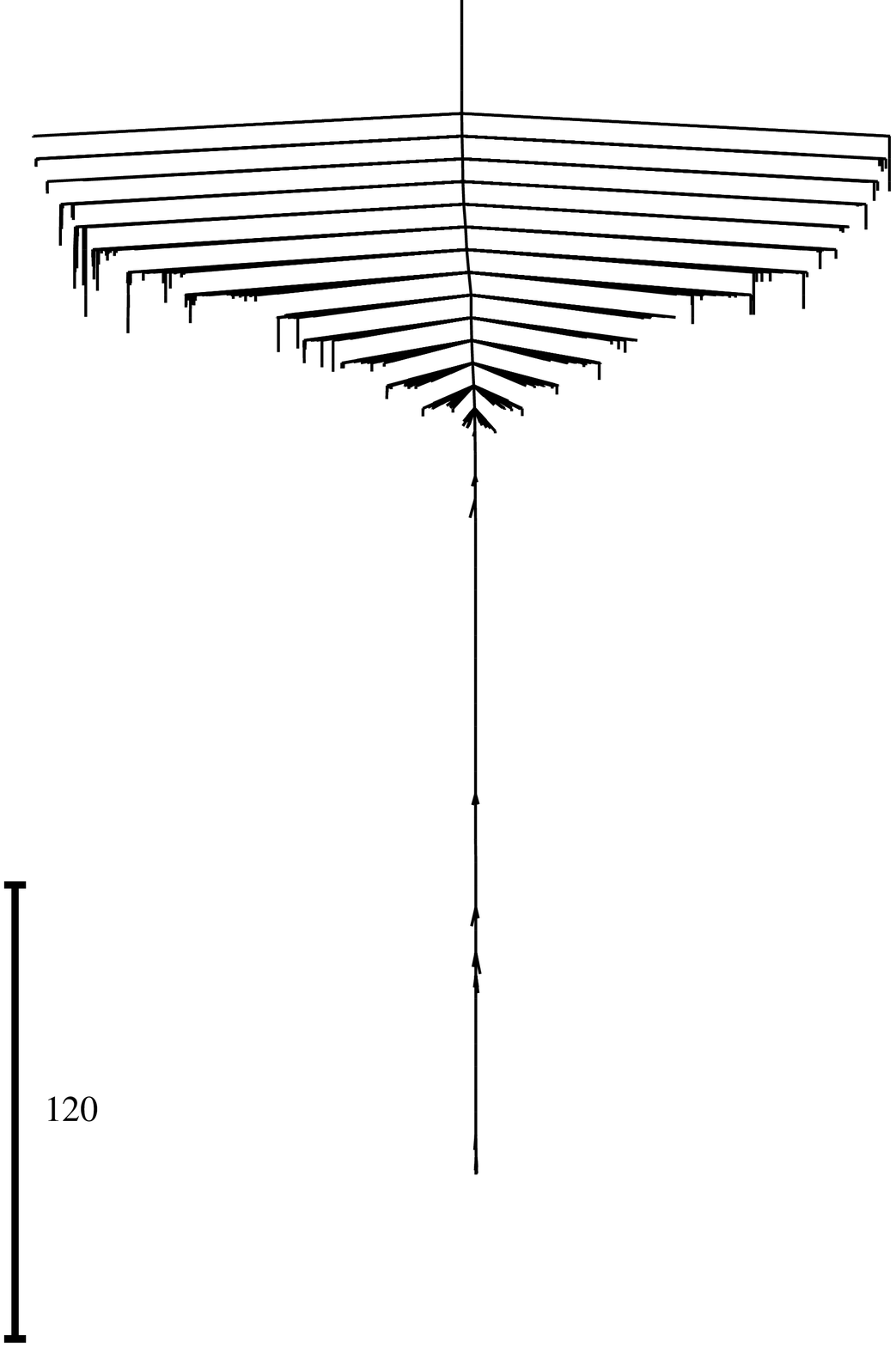}
        \caption{Armv1 ($1:2$)}
    \end{subfigure}%
    \begin{subfigure}{0.25\textwidth}
        \centering
        \includegraphics[width=0.95\textwidth]{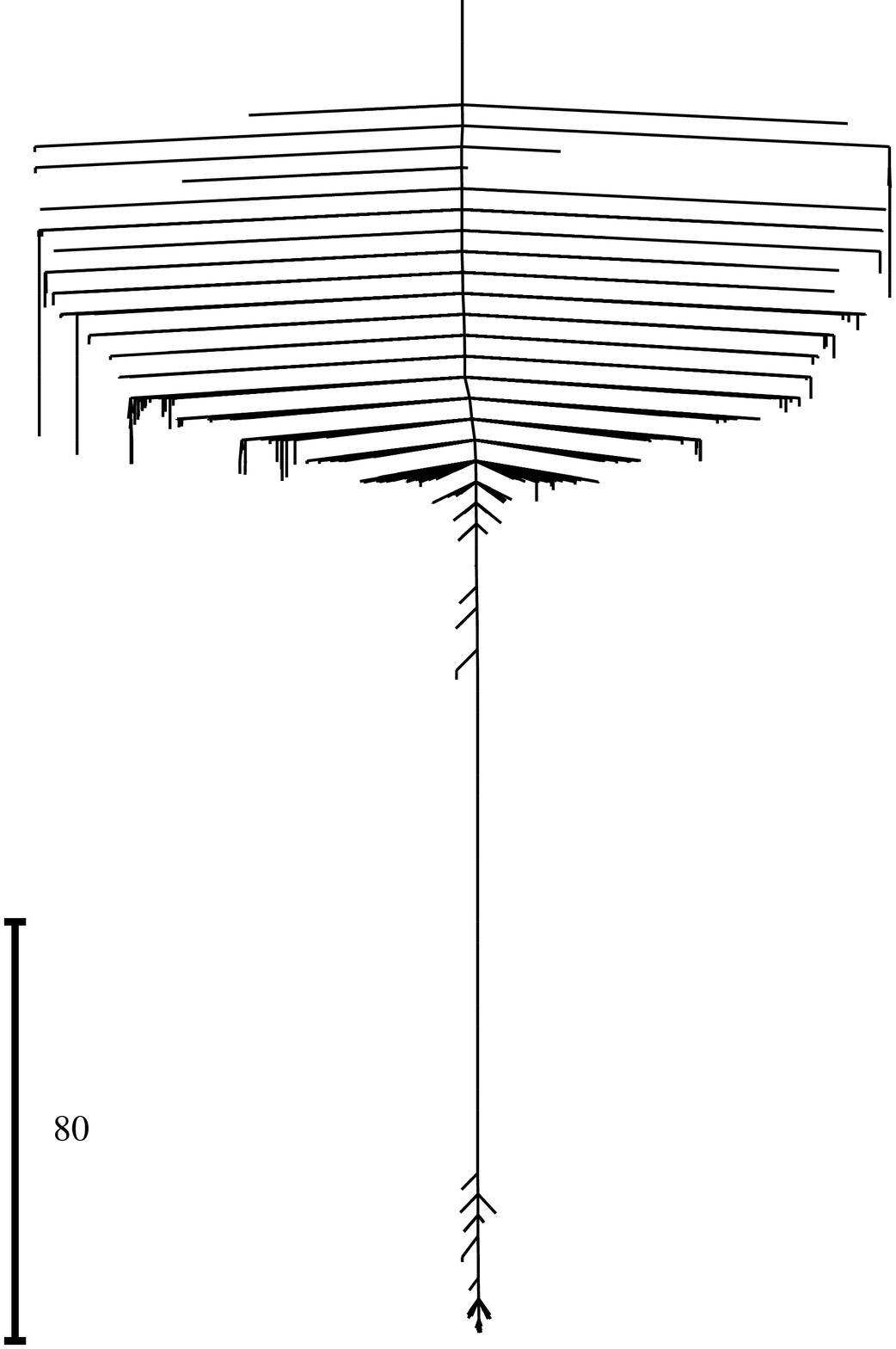}
        \caption{\small Shiv1 ($19:58$)}
    \end{subfigure}%
    \begin{subfigure}{0.25\textwidth}
        \centering
        \includegraphics[width=0.95\textwidth]{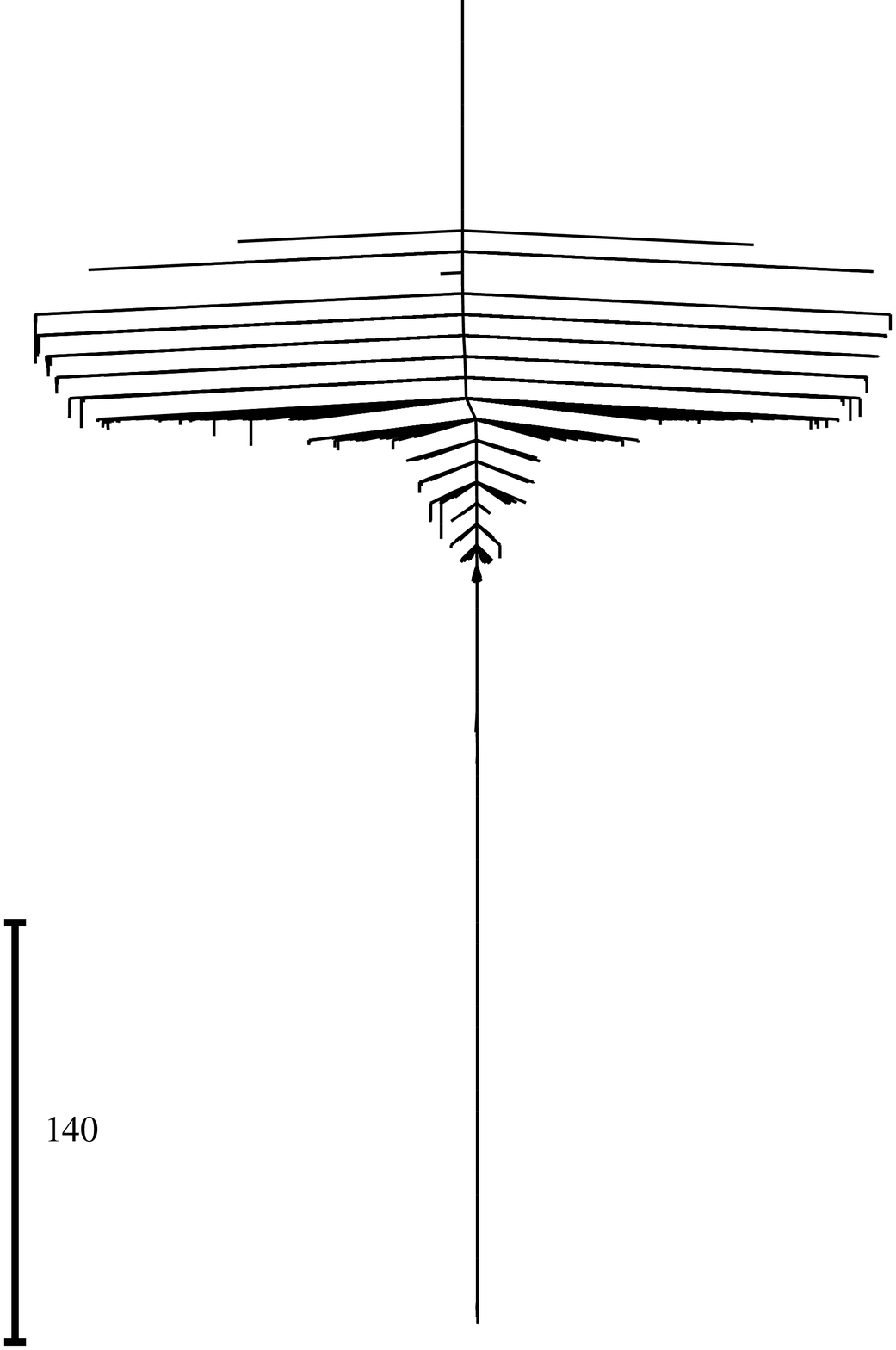}
        \caption{Gor ($31:150$)}
    \end{subfigure}\\
    \caption[$K$-means landscapes for varying samples per class ratios]{$K$-means landscapes for gene expression datasets with different samples per class ratios, all with two clusters.
The scale bar denotes the cost function value.}
    \label{fig:trees_vary_ratio}
\end{figure}

The landscapes remain single-funnelled at all ratios, as shown in Figure~\ref{fig:trees_vary_ratio}. The landscape organisation is remarkably preserved throughout the datasets. The increased samples per class ratio leads to fewer minima similar in cost function to the global minimum. This reduction in minima can be explained by the smaller cluster supporting fewer minima when a cluster centre is inside it. Hence, the samples per class ratio does not play an important role in determining the overall $K$-means landscape structure, but it can lead to a reduced number of low-valued clustering solutions. Therefore, location of the global minimum remains of comparable difficulty for optimisation algorithms, but it is increasingly important to find the global minimum as it becomes more unique.

\subsection{Importance of an appropriate $K$}

There are two main facets of clustering gene expression data: discerning the appropriate number of cancer subtypes, and accurate separation of cancer subtypes at a given number of clusters. In the previous sections we computed the $K$-means cost function topography for a known clinical number of clusters and rationalised $K$-means performance in distinguishing cancer subtypes and locating good $K$-means solutions. Here, we consider the determination of an appropriate number of cancer subtypes. Hence, the performance of $K$-means within a range of $K$ is important to identify an appropriate subdivision of cancerous cell lines into distinct subtypes.

\begin{figure}[h!]
    \centering
    \begin{subfigure}{0.32\textwidth}
        \centering
        \includegraphics[width=1.0\textwidth]{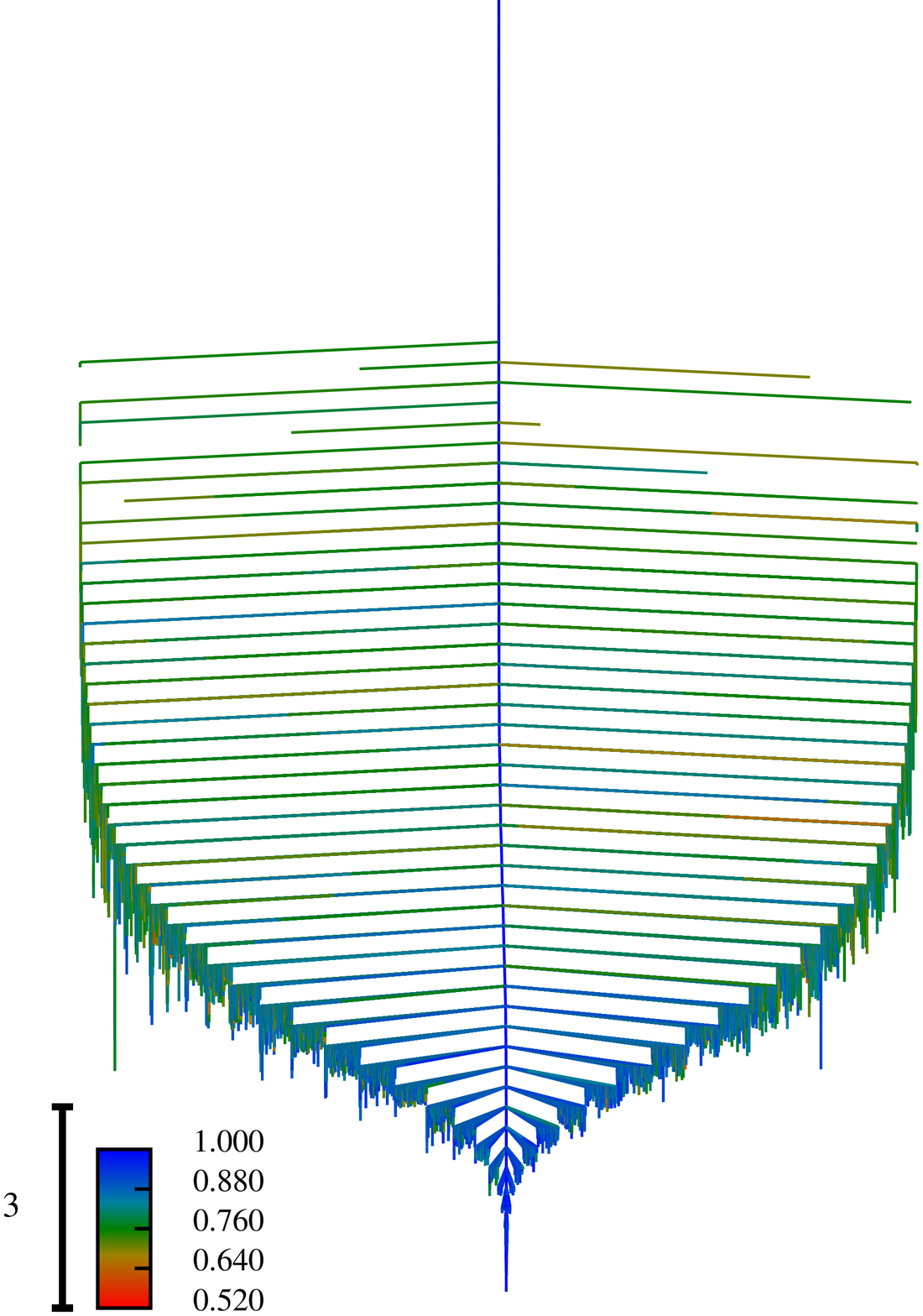}
        \caption{$K=2$}
    \end{subfigure}
    \begin{subfigure}{0.32\textwidth}
        \centering
        \includegraphics[width=1.0\textwidth]{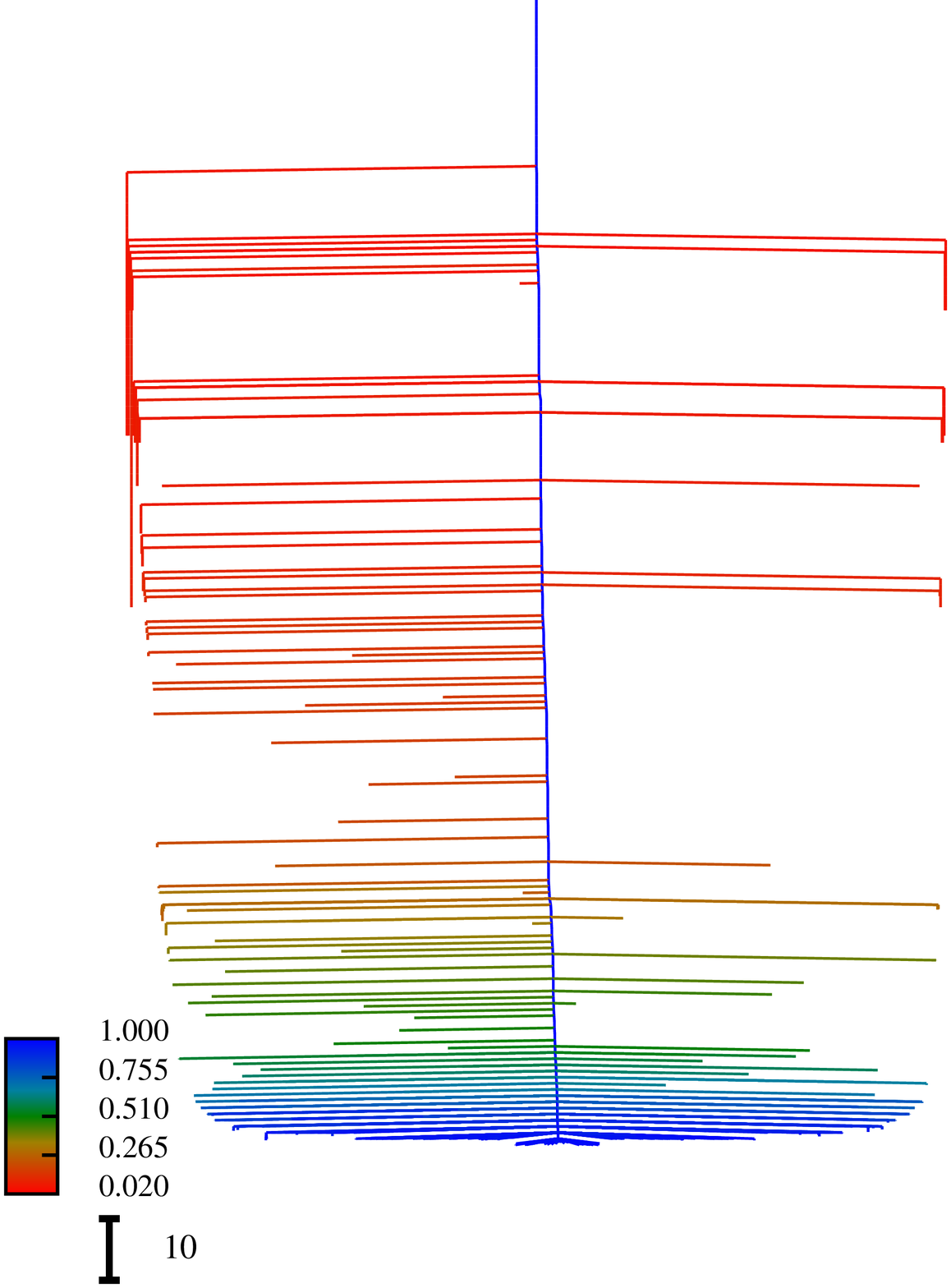}
        \caption{$K=3$}
    \end{subfigure}
    \begin{subfigure}{0.32\textwidth}
        \centering
        \includegraphics[width=1.0\textwidth]{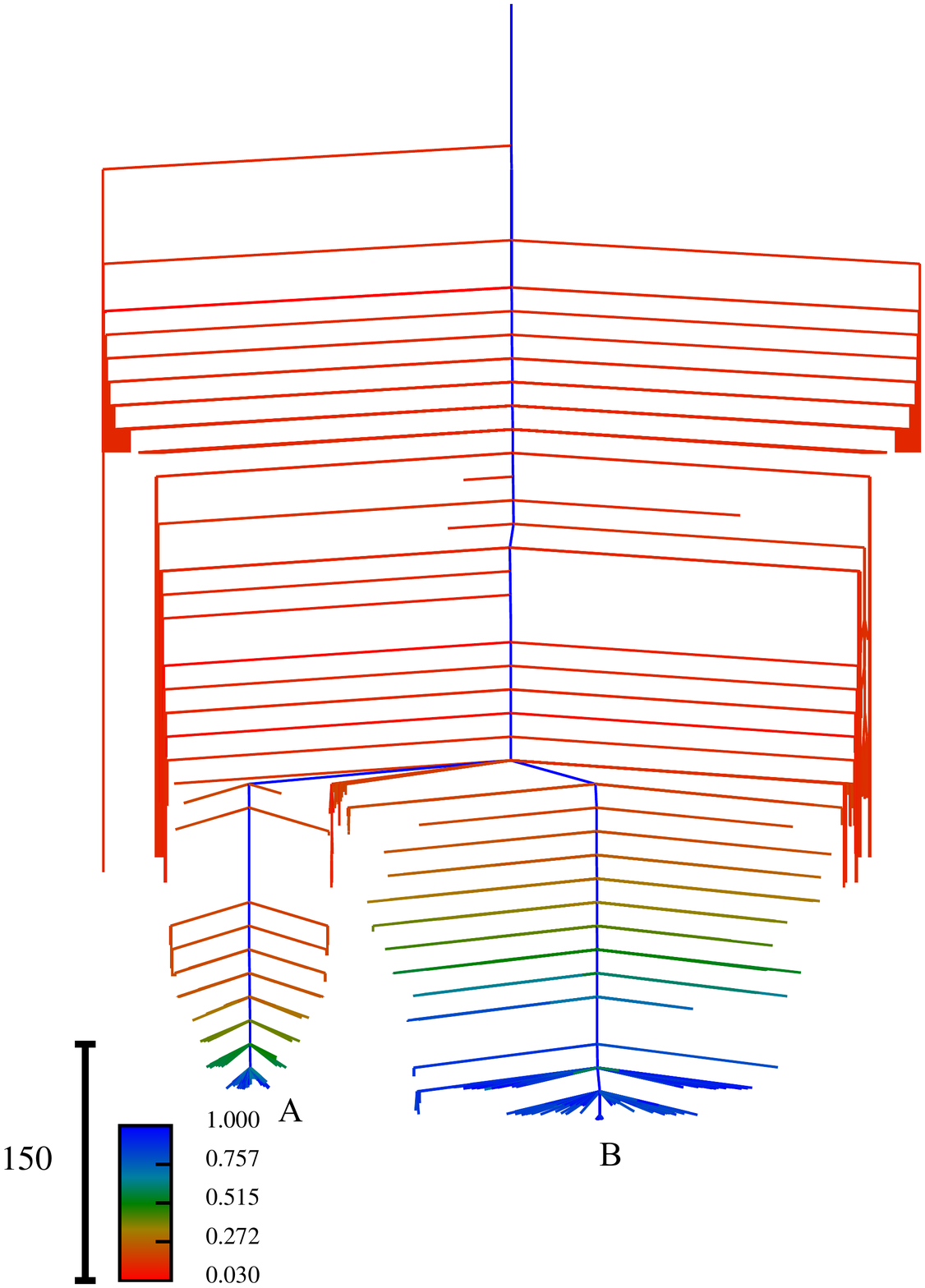}
        \caption{$K=4$}
    \end{subfigure}\\
    \begin{subfigure}{0.32\textwidth}
        \centering
        \includegraphics[width=1.0\textwidth]{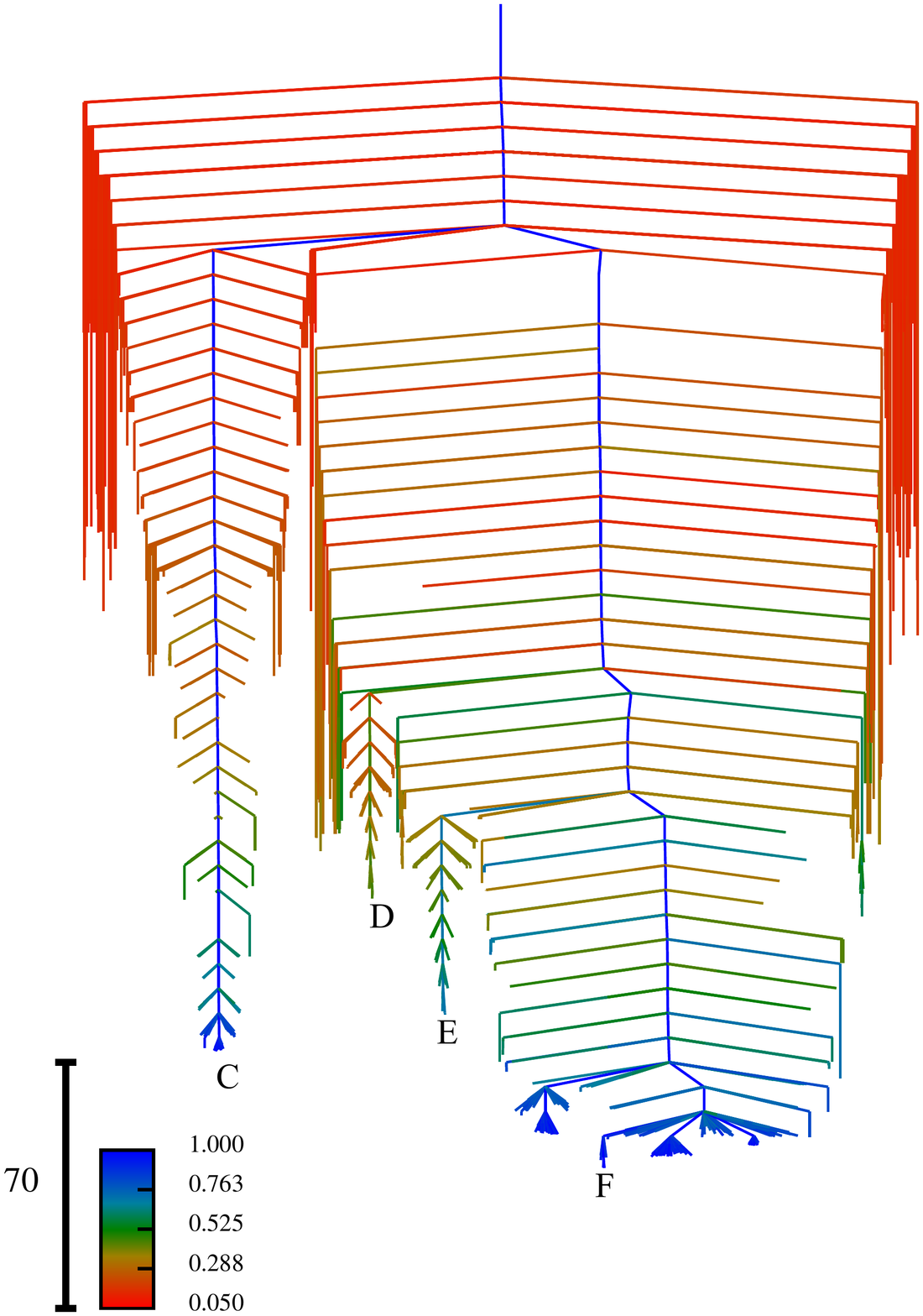}
        \caption{$K=5$}
    \end{subfigure}
    \begin{subfigure}{0.32\textwidth}
        \centering
        \includegraphics[width=1.0\textwidth]{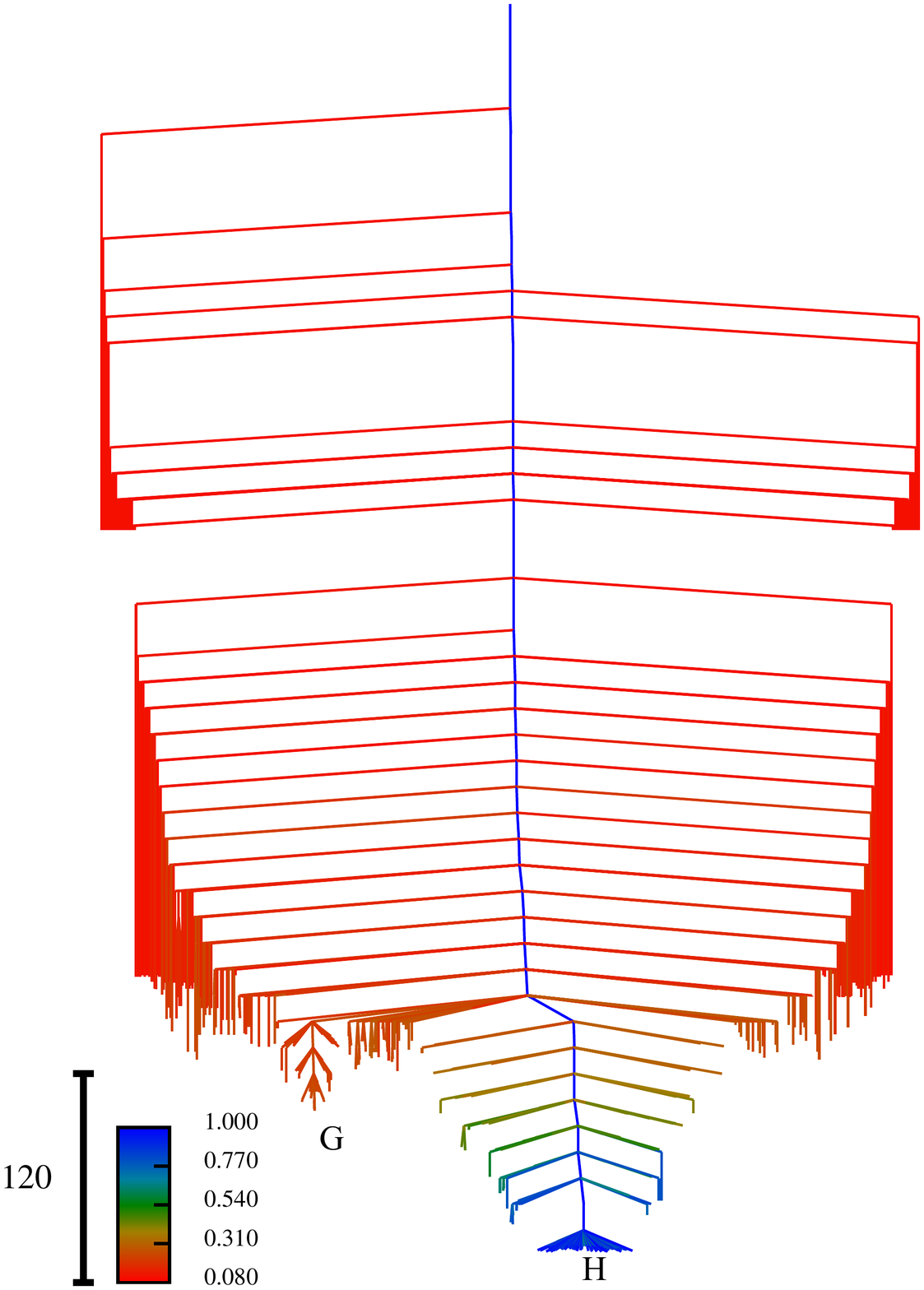}
        \caption{$K=6$}
    \end{subfigure}\\
    \caption[$K$-means landscapes for the Yeov2 dataset with different $K$]{$K$-means landscapes for the Yeov2 dataset with different $K$.
The minima are coloured by their ARIs relative to the global minimum clustering solution.
The range of ARIs is shown in the colour bar. 
The scale bar denotes the cost function value.
For $K=4,5,6$, minima at the bottom of the subfunnels are labelled alphabetically.}
    \label{fig:trees_yeo}
\end{figure}

The Yeoh dataset has two alternative clinical assignments, which differ in the number of cancer subtypes. Yeov1 separates leukaemia samples into two classes denoted T-ALL and B-ALL. Yeov2 further subdivides the B-ALL samples into five distinct classes (BCR-ABL, E2A-PBX1, Hyperdip$>$50, MLL and TEL-AML), increasing the total number of clusters to six. Here, we vary $K$ from two to six to demonstrate the impact of $K$ on the landscape structure, which is shown in Figure~\ref{fig:trees_yeo}.

From $K=2$ to $K=5$, the solution space becomes more multi-funnelled as $K$ deviates from the true number of clusters ($K=2$). At $K=3$, the landscape remains single funnelled, but the high cost function region is now very different from
the region at low $J$. For $K=4$ and $K=5$, the single funnel structure is lost and there are two or more competing funnels. At $K=6$, the single funnel structure is largely restored, although a small subfunnel still exists. Generally, we observe that when $K$ matches the true number of classes ($K=2$ or $K=6$ in this example), the $K$-means landscape has a largely single-funnel structure to the global minimum. In contrast, when the number of clusters is not assigned correctly, the landscape becomes more structured, with an increasing number of subfunnels that will impede location of the global minimum. 

\begin{figure}[h!]
    \centering
    \includegraphics[width=1.0\textwidth]{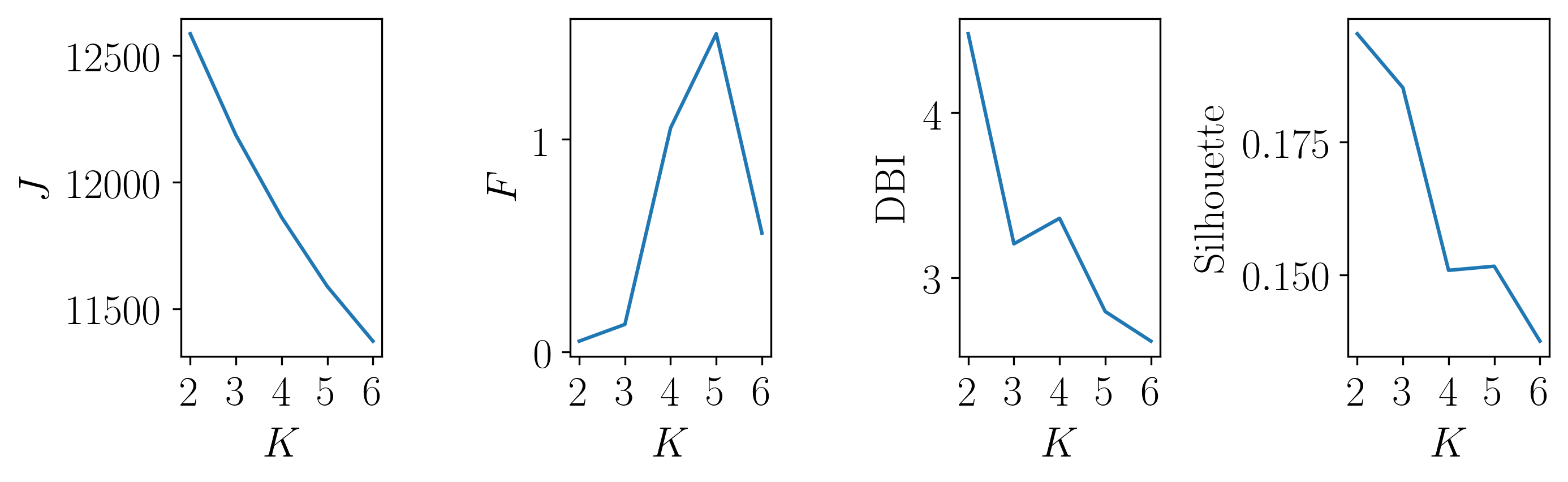}
     \caption{Several common metrics for determining the appropriate $K$ applied to the Yeohv2 gene expression dataset. From left to right the metrics are cost function value, frustration metric,\cite{deSouza2017} Davies-Bouldin index\cite{Davies1979} and the silhouette score.\cite{Rousseeuw1987}}
    \label{fig:cluster_selection}
\end{figure}

We can quantify the single-funnelled character of the $K$-means landscapes using a frustration metric,\cite{deSouza2017} which is described in detail in Sec.~SII. Small values of the frustration metric correspond to single-funnelled landscapes, and the metric increases with larger barriers to lower-valued minima and more subfunnels in the landscape. We compute the frustration metric for the range of $K$ and compare with other popular metrics for selecting the appropriate number of clusters in Fig.~\ref{fig:cluster_selection}.

In this example, the frustration metric provides a good measure by which to identify the clinical number of cancer subtypes. Choosing an appropriate number of clusters for datasets with few datapoints and relatively uniform distances due to the high dimensionality is challenging, and methods based on within and between-cluster variance do not work very well. The silhouette score, Davies-Bouldin index and elbow plot do not exhibit a clear preference for the clinical number of clusters, $K=2$ and $K=6$. In constrast, the frustration metric, which clearly captures the single-funnelled nature of the landscapes, exhibits a sharp decrease at $K=6$ after a minimum for $K=2$. Therefore, by leveraging a metric that considers the whole solution space, rather than a single minimum, it may be possible to deduce the appropriate cluster number, even in gene expression datasets that are challenging for standard metrics based on the cluster variances.

For landscapes with significant subfunnels ($K=4$ and $K=5$), the minima in two different subfunnels may have significantly different accuracy at the same cost function level. Landscapes provide a natural way of separating distinct types of of clustering solution that may not be possible with other methods. Minima belonging to different subfunnels correspond to different types of solutions, and we can identify the specific data points that mediate the change between types of clustering solutions. A detailed analysis of the different cluster solution types and their compositions is given in Sec.~SIII.

\section{Conclusions}

In this study, we investigated the $K$-means landscapes of gene expression datasets. We probed the effects of several important dataset properties on the solution landscapes; the clinical number of clusters, the number of features, and the cluster distribution. We observe that variations in these features do not significantly modify the $K$-means landscape topography. The landscapes remain largely funnelled in all datasets when the appropriate cluster number is used. This structure indicates that the location of the global minimum will remain a problem of comparable complexity for $K$-means optimisation algorithms.\cite{Franti2000, Yao2016, Das2018, Xie2019} However, we note that a more even cluster distribution and increasing features leads to more low-valued minima resembling the global minimum, so obtaining the true global minimum is probably less important.

In contrast, varying the number of clusters used to represent a given dataset has a marked effect on landscape topography. As $K$ increases the landscape becomes more structured and contains additional subfunnels, but it still remains largely single funnelled if the chosen $K$ matches the clinically-assigned number of clusters. Deviation from the true $K$ may lead to the formation of sub-funnels, which increase the difficulty in locating the global minimum. We observe that the subfunnels indicate different types of clustering solution and, consequently, the barriers provide an alternative measure of clustering distinctness, which does not rely on cluster labels or Euclidean distance. 

Finally, we apply a global frustration metric to quantify the landscape structure, and show that this method can highlight the clinical number of clusters. Determining the appropriate number of clusters is challenging in these datasets with few points in high dimensions, but a global analysis of the whole solution space still allows the determination of clinical cluster number, whereas standard metrics based on cluster variance do not. Therefore, we predict that the landscape structure can be a good indicator for the choice of $K$ for gene expression datasets, and the landscape will be single funnelled if the correct $K$ is chosen. This prediction needs to be evaluated 
in future work for additional gene expression datasets with multiple correct labellings and differing numbers of classes.

In line with previous studies, we observe poor accuracy in reproducing clinical assignments with $K$-means for many of these datasets, which contain only a small number of high-dimensional samples. Moreover, in many cases there is little correlation between accuracy and cost function, which suggests that locating low-valued solutions of the $K$-means landscape may not produce the most accurate clusterings. Therefore, it is challenging to select an appropriate clustering solution, and it is clear that alternative metrics should be considered when selecting $K$-means, and ensemble methods would aid performance.

In summary, this work presents the first examples of $K$-means landscapes for cancer gene expression datasets. Such landscapes enable us to understand how dataset properties affect the $K$-means performance in identifying cancer subclasses, which is crucial for cancer diagnosis. Moreover, the landscapes provide a possible route to determining the appropriate number of cancer subtypes for novel datasets.

\section{Supporting Information}

The supporting information includes a detailed discussion of the choice of appropriate parameters for the surrogate function in location of transition states, a mathematical description of the frustration metric, and a discussion of cluster composition for the Yeo dataset with varying cluster number. The code used to generate all results in this work can be freely accessed at:
git@gitlab.developers.cam.ac.uk:ch/wales/softwarewales.git.

\section{Acknowledgments}

LD would like to acknowledge the EPSRC for funding under a knowledge transfer fellowship. DJW gratefully acknowledges financial support from the EPSRC and an International Chair at the Interdisciplinary Institute for Artificial Intelligence at 3iA Cote d’Azur, supported by the French government, with reference number ANR-19-P3IA-0002, which has provided interactions that furthered
the present research project.

\bibliographystyle{unsrt}
\bibliography{./KMeans_GeneExpression_References.bib}

\end{document}



\title[]{Supporting Information - Archetypal solution spaces for clustering gene expression datasets in identification of cancer subtypes} 

\maketitle

\section{Optimising the surrogate function parameters}

The accuracy of transition state location is important when constructing $K$-means landscapes.
Gene expression datasets contain data points of significantly higher
dimensionality than previous applications to the $Iris$ dataset in
Ref.~\citen{DicksW22}. 
We therefore reevaluated the surrogate function parameters ($\alpha$ and
$\sigma$) used to locate transition states. We tested different combinations of
$\alpha$ and $\sigma$ in $F_+$, and measured the error in transition state
location, which was here defined as the difference in cost function at the
MECP, evaluated for both $\bm{r}^{(1)}$ and $\bm{r}^{(2)}$.

In Ref.~\citen{DicksW22}, $\alpha=0.02$ and $\sigma=30$ were used for the \textit{Iris} dataset.
Here, we employed a grid search strategy for the parameter choices.
The values were chosen both on a linear scale ($\alpha=\{0.01,0.02,\ldots,
0.09\}$ and $\sigma=\{10,20,\ldots,90\}$) and on a $\log_2$ scale
($\alpha=\{0.0025,0.005,\ldots, 0.16\}$ and $\sigma=\{3.75,7.5,\ldots,240\}$).
For each combination of $\alpha$ and $\sigma$, we attempted ten connections
between different pairs of minima, and calculated the errors for all the transition
states.

\begin{figure}
    \centering
    \begin{subfigure}{0.45\textwidth}
        \centering
        \includegraphics[width=1.0\textwidth]{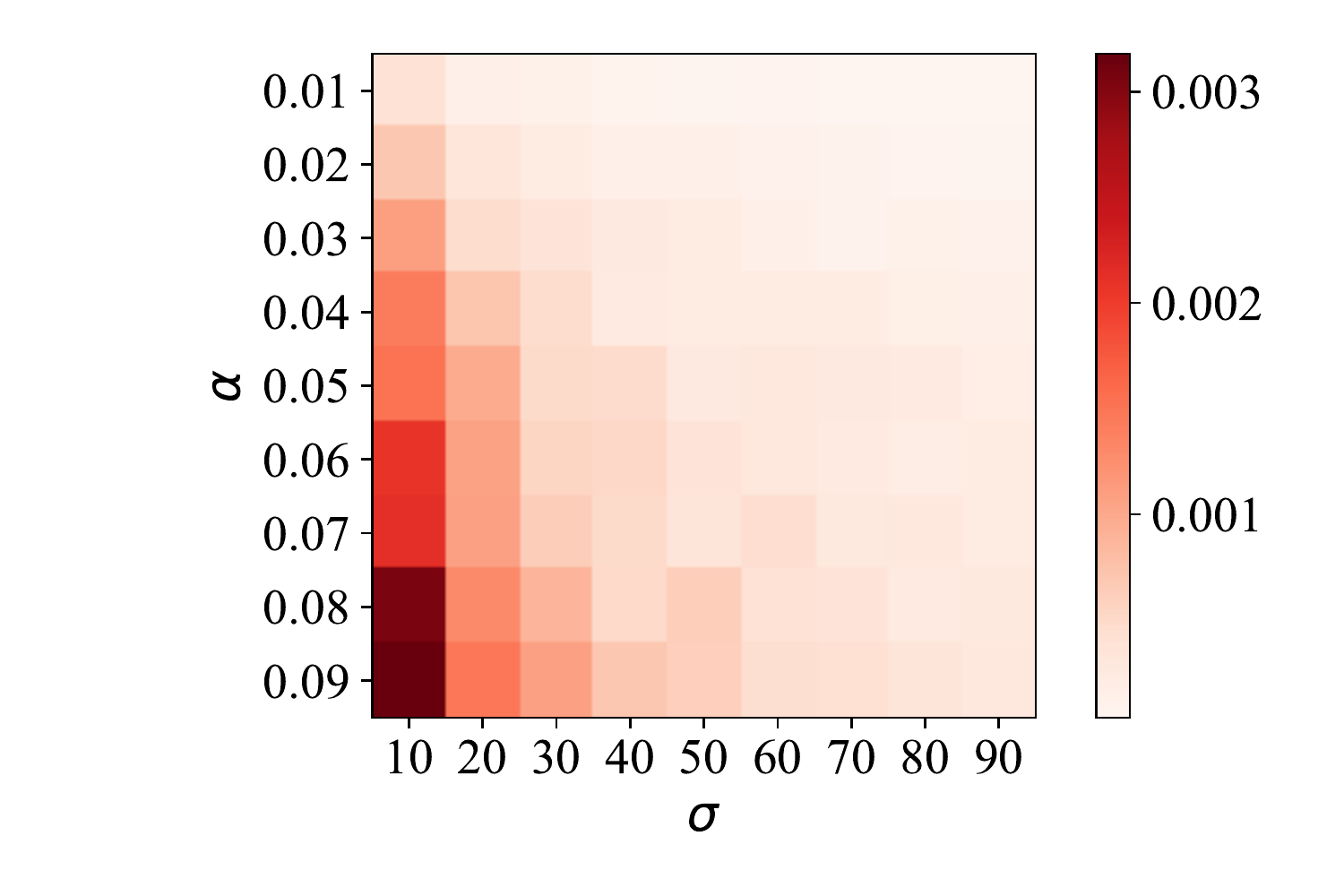}
        \caption{Nutv3, linear}
    \end{subfigure}
    \begin{subfigure}{0.45\textwidth}
        \centering
        \includegraphics[width=1.0\textwidth]{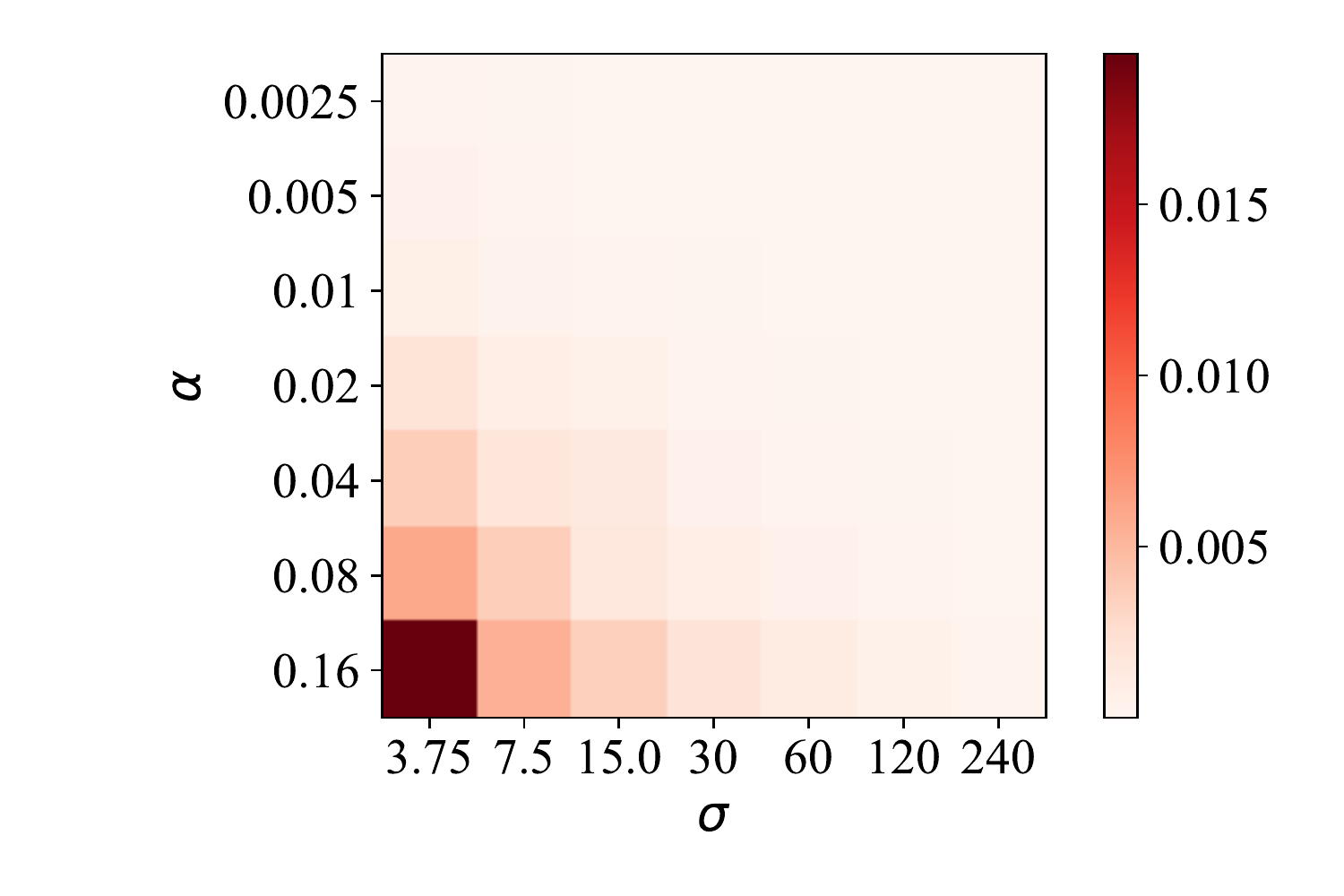}
        \caption{Nutv3, logarithmic}
    \end{subfigure}  
    \begin{subfigure}{0.45\textwidth}
        \centering
        \includegraphics[width=1.0\textwidth]{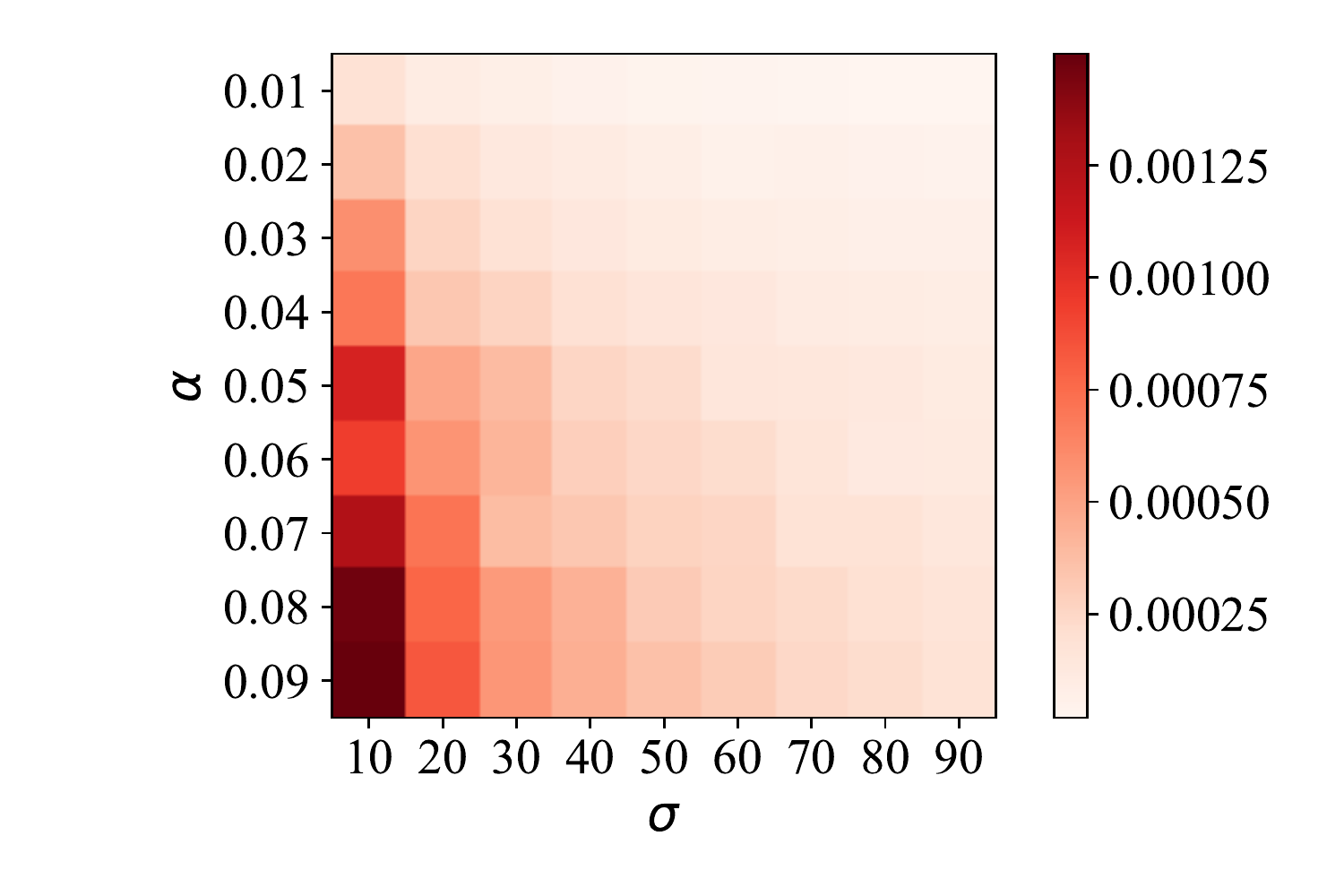}
        \caption{Dyr, linear}
    \end{subfigure}
    \begin{subfigure}{0.45\textwidth}
        \centering
        \includegraphics[width=1.0\textwidth]{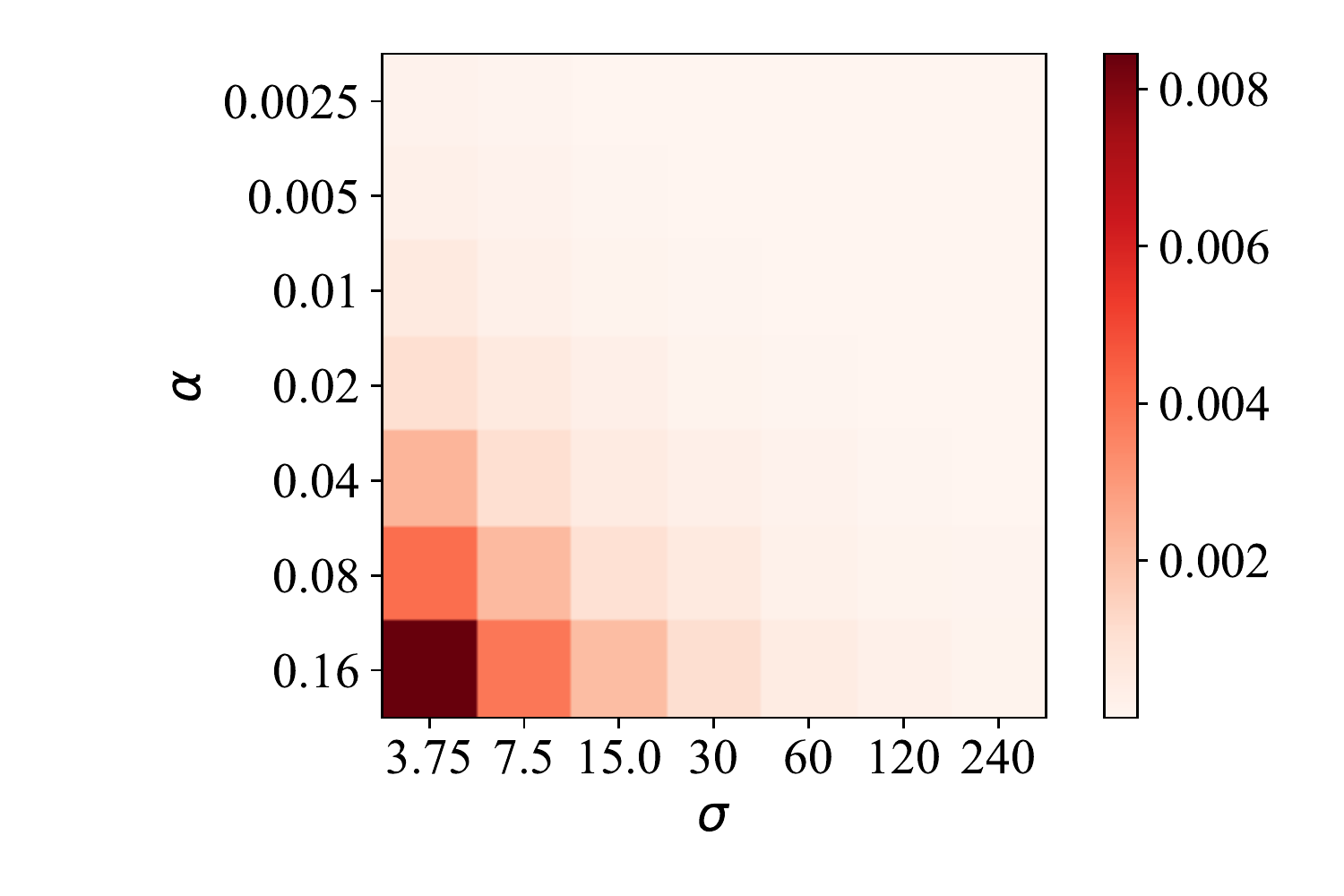}
        \caption{Dyr, logarithmic}
    \end{subfigure} 
    \begin{subfigure}{0.45\textwidth}
        \centering
        \includegraphics[width=1.0\textwidth]{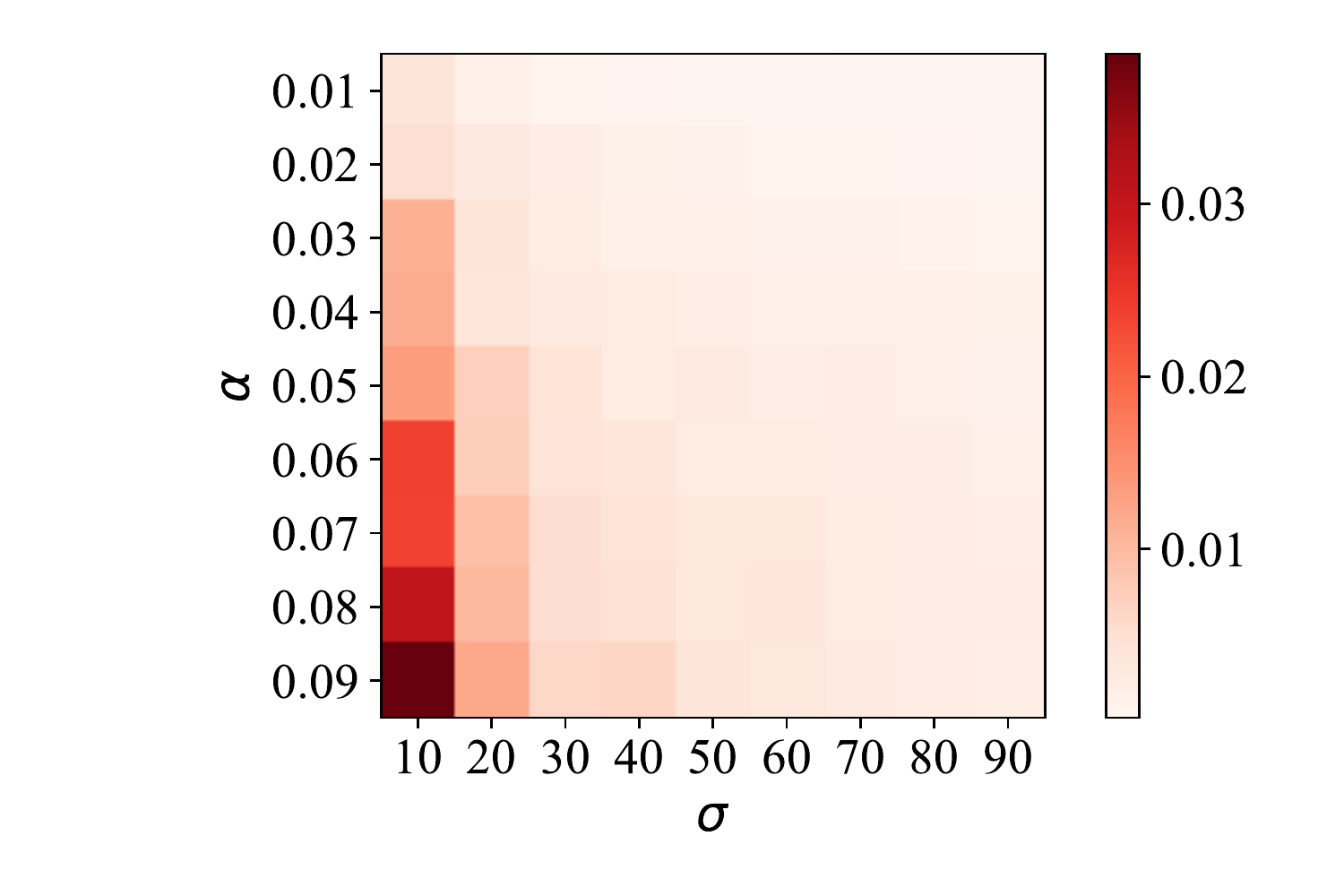}
        \caption{\textit{Iris}, linear}
    \end{subfigure}
    \begin{subfigure}{0.45\textwidth}
        \centering
        \includegraphics[width=1.0\textwidth]{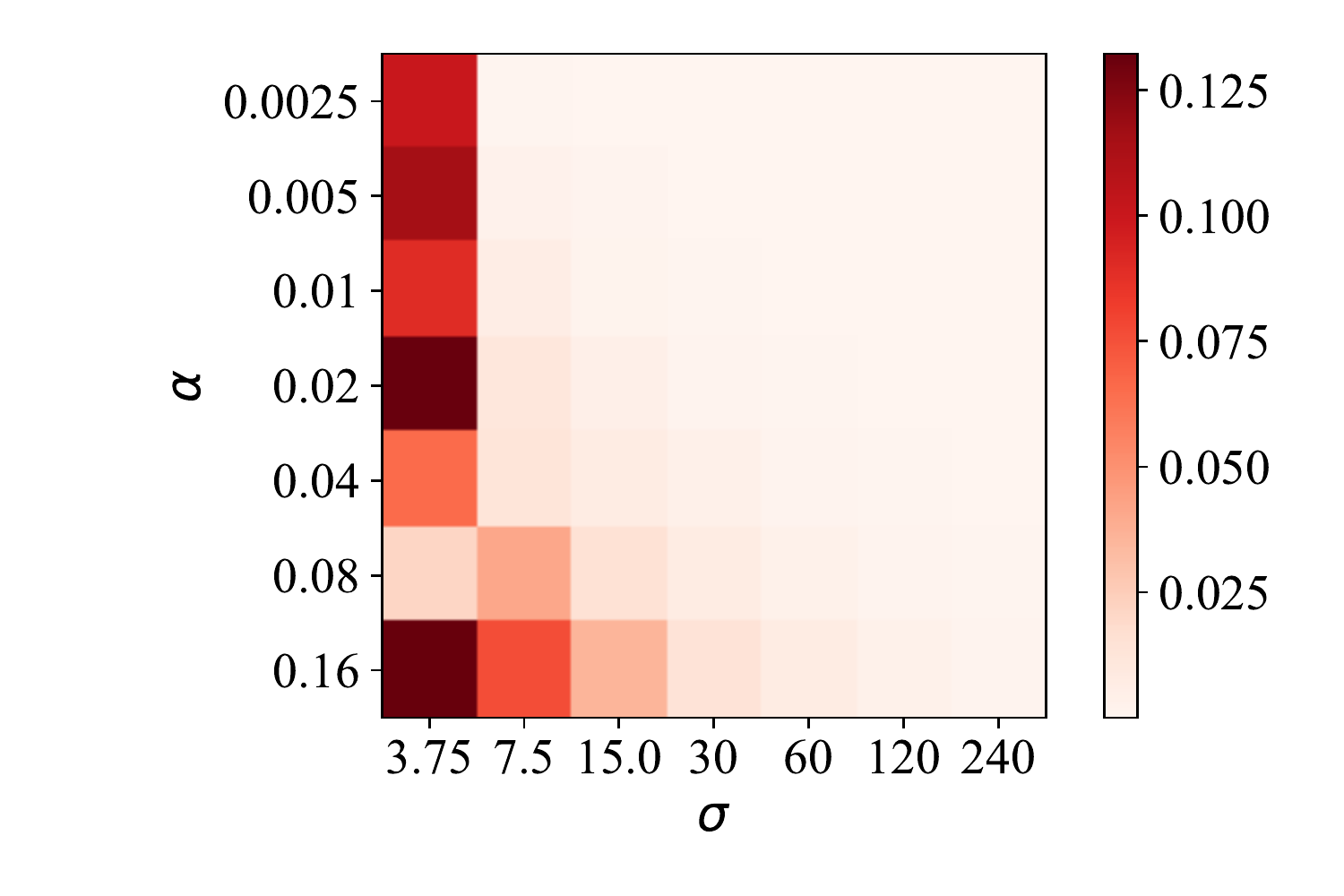}
        \caption{\textit{Iris}, logarithmic}
    \end{subfigure}
    \caption[Heatmaps of transition state errors]{Heatmaps of transition state errors for
Nutv3 ($K=2$), Dyr ($K=3$) and \textit{Iris} ($K=3$).
$\alpha$ and $\sigma$ are varied on both linear and logarithmic scales.}
    \label{fig:heatmap}
\end{figure}

\begin{figure}
    \centering
    \includegraphics[width=0.85\textwidth]{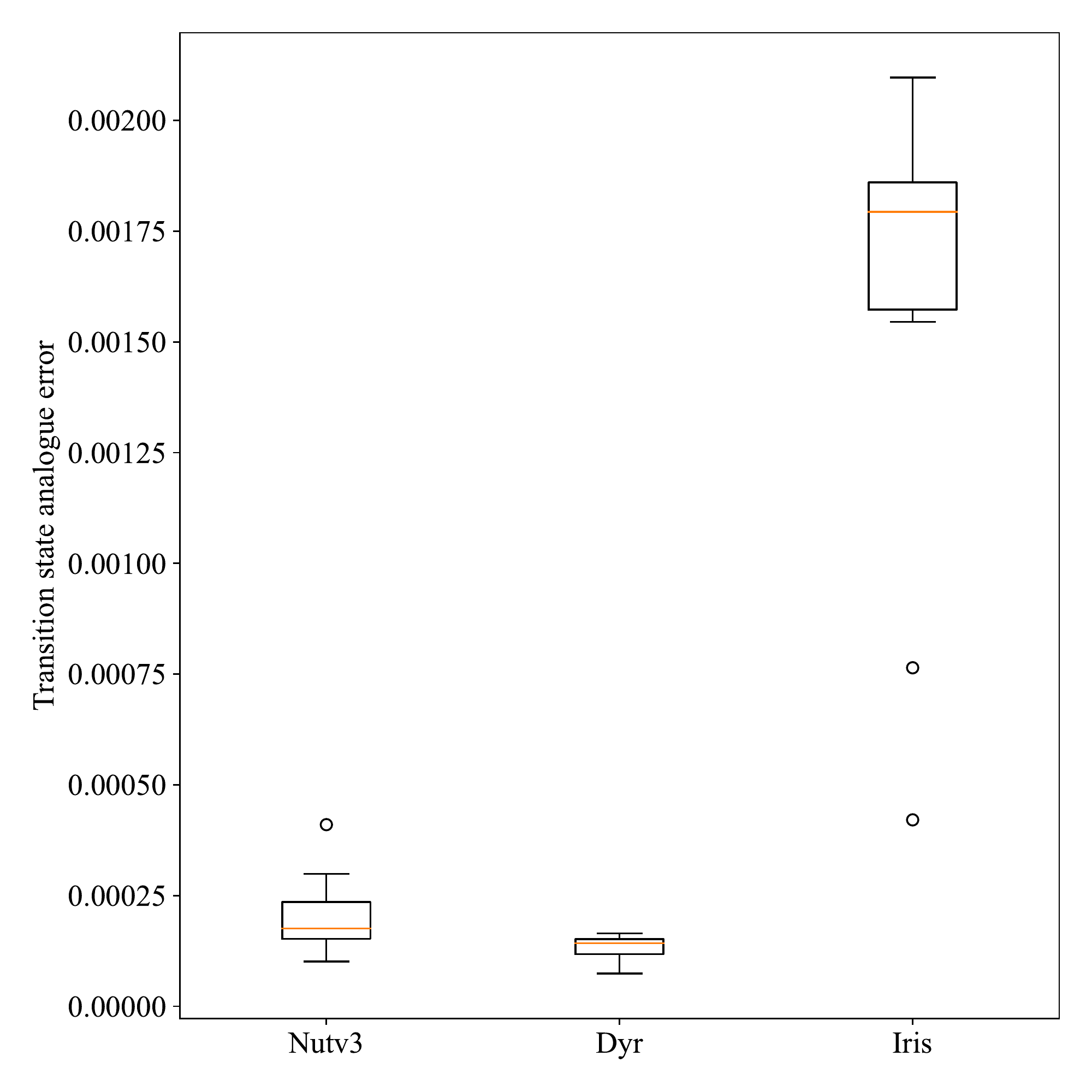}
    \caption[Box plots of transition state errors]{Box plots of transition state errors when using $\alpha=0.02$ and $\sigma=30$
for Nutv3 ($K=2$), Dyr ($K=3$) and \textit{Iris} ($K=3$).
The box extends from the first quartile to the third quartile of the data, with an orange line at the median.
The whiskers extend to 1.5 times the inter-quartile range. 
Values outside whiskers are shown in circles.
}
    \label{fig:boxplots}
\end{figure}

To illustrate how the error varies with parameter choices, heatmaps of errors
for the Nutv3, Dyr and \textit{Iris} datasets are shown in
Figure~\ref{fig:heatmap}.
We find that as $\alpha$ decreases, or as $\sigma$ increases, the error becomes smaller and the location of the transition state is more accurate.
However, making $\alpha$ too small will cause a discontinuity at the intersection seam, and making $\sigma$ too large will slow the minimisation.
The choice of $\alpha=0.02$ and $\sigma=30$ proposed in Ref.~\citen{DicksW22} appears to work well for all three datasets.
It achieves small errors, and further reduction in the error is insignificant if we decrease $\alpha$ or increase $\sigma$.
Figure~\ref{fig:boxplots} shows that, when $\alpha=0.02$ and $\sigma=30$,
errors for two gene expression datasets are similar [$\mathcal{O}(10^{-4})$],
and are around ten times smaller than those for the \textit{Iris} dataset.
Therefore, the choice of $\alpha=0.02$ and $\sigma=30$ is successful for
datasets with different properties and can accurately locate transition states
for gene expression datasets.

\clearpage

\section{Frustration metric} \label{frustration}

There are various measures that have been derived to quantify the degree of solution space funnelling. The single or multi-funnelled character of a landscape can be related to the frustration,\cite{Bryngelson1995, Onuchic1997} which describes the difficulty in attaining the global minimum of a system. Single-funnelled landscapes typically have little frustration and efficiently reach the unique global minimum due to the absence of competing morphologies with high barriers between them. In contrast, highly frustrated systems relax to the global minimum much more slowly, and are associated with landscapes that have significant barriers between numerous low-lying minima.

The Shannon entropy largely correlates with frustration,\cite{deSouza2017} and allows us to quantify the single-funnelled character of a given $K$-means landscape. The Shannon entropy,\cite{Shannon1948} $s(T)$, is defined as
\begin{equation}
s(T) = - \sum_{\alpha} p_{\alpha}^{\mathrm{eq}}(T) \mathrm{ln} p_{\alpha}^{\mathrm{eq}}(T),
\end{equation}
where $T$ is the temperature, and $p_{\alpha}^{\mathrm{eq}}$ is the population of minimum $\alpha$ at equilibrium. The equilibrium occupation probability, assuming a harmonic vibrational density of states,\cite{Wales2003} is defined as
\begin{equation} \label{Populations}
p^{\mathrm{eq}}_\alpha(T) = \frac{\mathrm{e}^{-J_{\alpha}/k_{\mathrm{B}}T}/\bar{\nu}_{\alpha}^{\kappa}}{\sum_{\gamma} \mathrm{e}^{-J_{\gamma}/k_{\mathrm{B}}T}/\bar{\nu}_{\gamma}^{\kappa}}.
\end{equation}
$k_B$ is the Boltzmann constant, $J_{\alpha}$ is the cost function value of minimum $\alpha$, and $\kappa$ is the number of non-zero eigenvalues of the Hessian matrix, in this case $\kappa = K \times N_f$. $\bar{\nu}_{\alpha}$ is the geometric normal mode vibrational frequency. For gene expression landscapes the normal mode analysis simply employs unit masses to derive a vibrational density of states. $k_B T$ is in the same units as $J_\alpha$ to produce a dimensionless ratio.

\clearpage

\section{Cluster composition} \label{cluster_composition}

\begin{figure}[h!]
    \centering
    \begin{subfigure}{0.85\textwidth}
        \centering
        \includegraphics[width=1.0\textwidth]{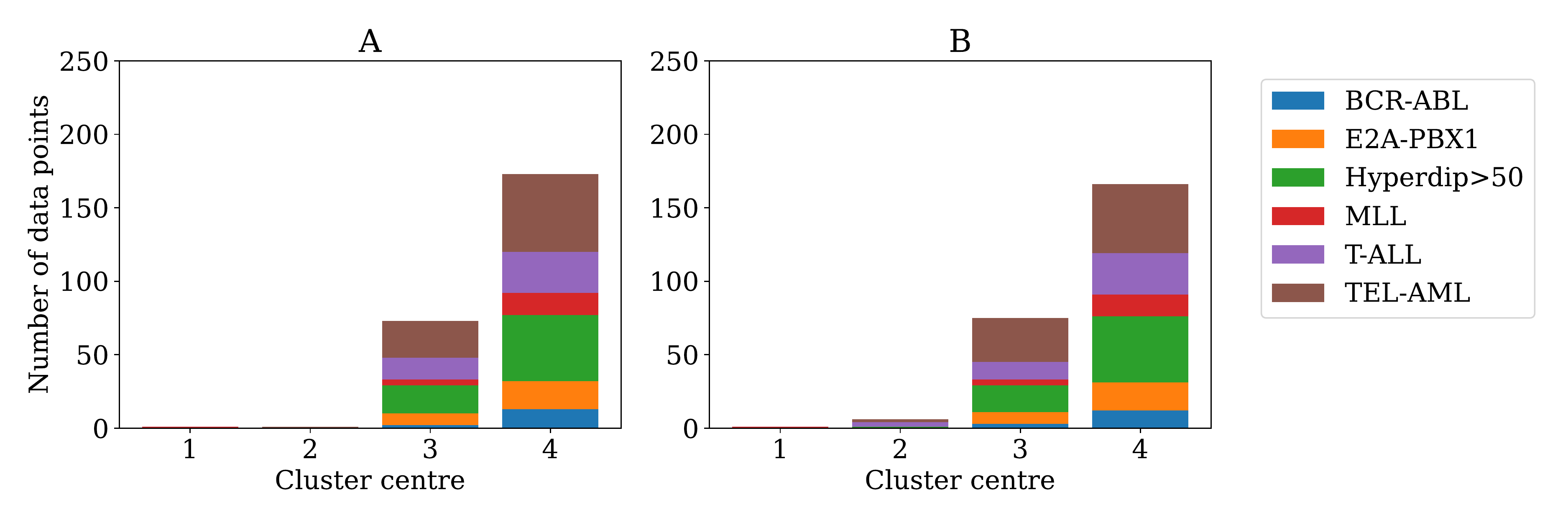}
        \caption{$K=4$}
    \end{subfigure}
    \begin{subfigure}{0.85\textwidth}
        \centering
        \includegraphics[width=1.0\textwidth]{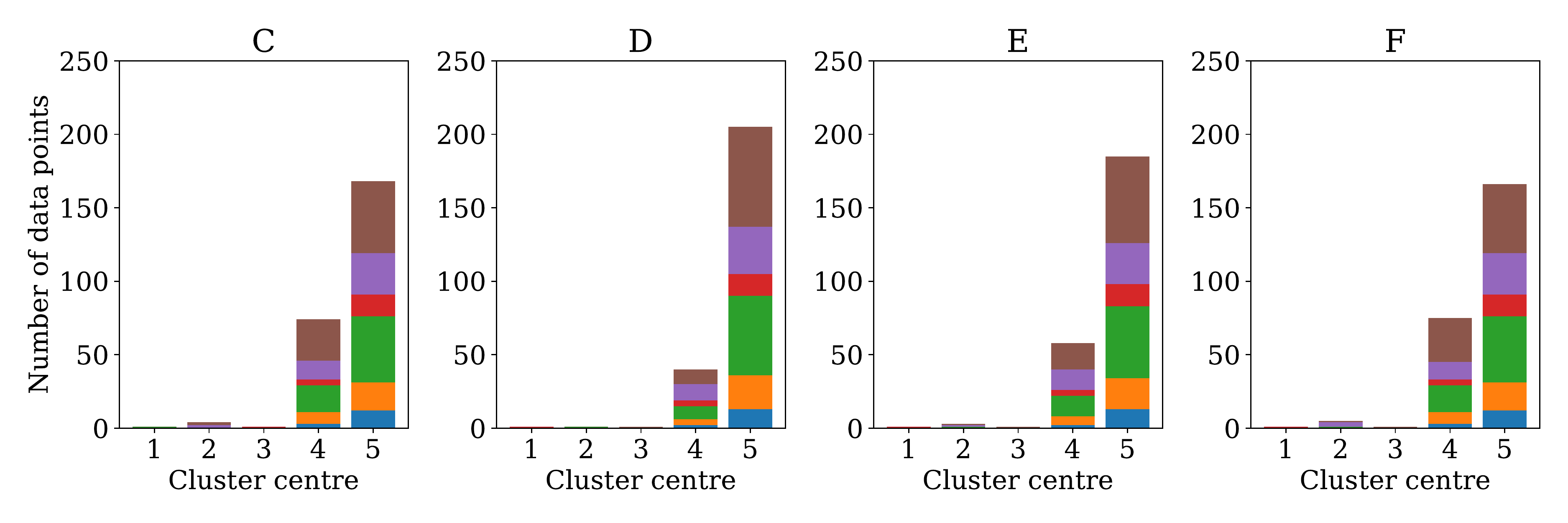}
        \caption{$K=5$}
    \end{subfigure}
    \begin{subfigure}{0.85\textwidth}
        \centering
        \includegraphics[width=1.0\textwidth]{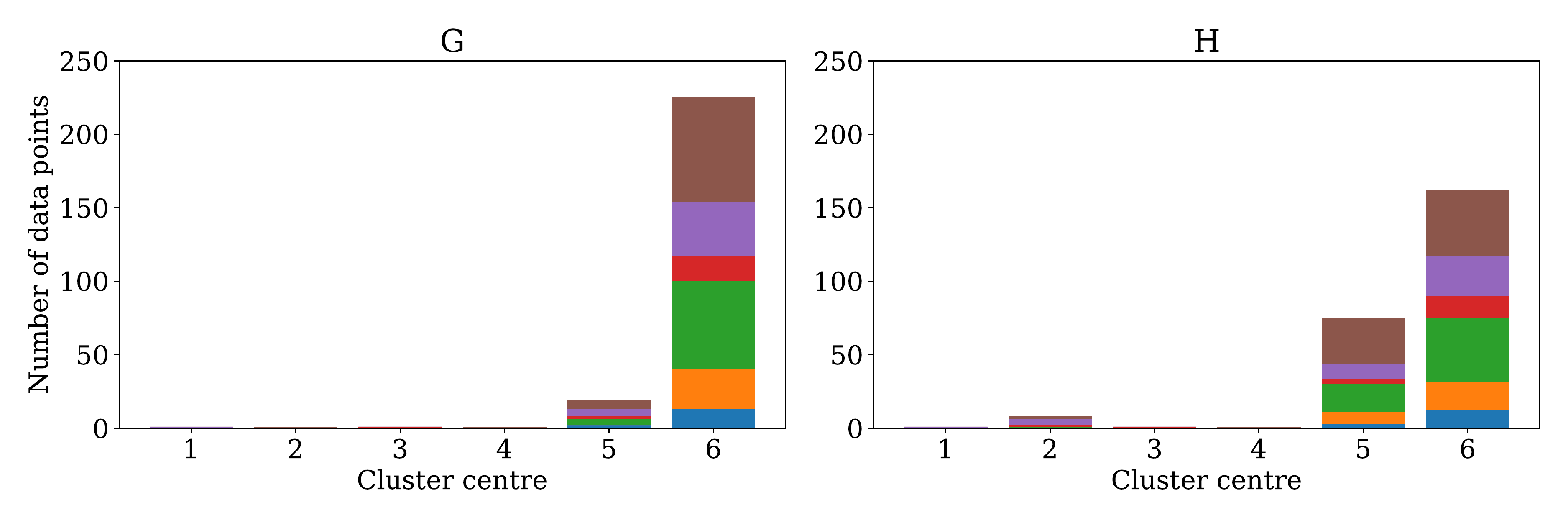}
        \caption{$K=6$}
    \end{subfigure}
    \caption[Composition of clustering solutions in the Yeov2 dataset]{Composition of selected clustering solutions computed for the Yeov2 dataset with $K=4,5$ and $6$. The composition is calculated with reference to the six clinical cancer subtypes shown in (a). Minima from each subfunnel were selected, and their positions on the landscape are labelled in Figure~7.}
    \label{fig:compos_yeo}
\end{figure}

The composition of these different clusterings, shown in Figure~\ref{fig:compos_yeo}, can be used to understand the subfunnels, each of which contains a distinct type of clustering solution. For $K=6$, we observe that neither the global minimum, nor the lowest minimum within the subfunnel, can separate six cancer subclasses accurately. Moreover, the cluster membership is very uneven, with almost all data points localised in two clusters. The global minimum has less disparity in the cluster sizes, leading to greater accuracy. The uneven membership could be explained by the plausible clinical assignment with only two clusters, which leads to preferential separation into two large $K$-means clusters, with the remaining clusters containing a small number of data points that could be considered outliers. For $K=4$ and $K=5$, there remain only two significantly populated clusters in all funnels, with similar composition to those seen for $K=6$, and a poor separation of true classes. These solutions indicate that additional clusters from $K=4$ to $K=6$ are best accommodated by assignment to a small number of outlier data points, preserving two large clusters.

For $K=4$, the main difference between Minimum A and Minimum B results from the exchange of one BCR-ABL data point and six TEL-AML data points, between clusters 3 and 4. For $K=5$, the principal difference between Minimum C and Minimum F is associated with the assignment of two TEL-AML data points to either cluster 5 or 4, respectively. These variations demonstrate the importance of these points in determining the resultant clustering, and that they could probably be considered outliers.

\clearpage

\bibliographystyle{unsrt}
\bibliography{./KMeans_GeneExpression_References.bib}